\newcommand*{\ATLASLATEXPATH}{}
\documentclass[UKenglish,texlive=2018,atlasstyle=true,paper=a4,cernpreprint,txfonts]{\ATLASLATEXPATH atlasdoc}
 
\usepackage[biblatex=true,backend=biber,subcaption=true,default]{\ATLASLATEXPATH atlaspackage}
\usepackage{\ATLASLATEXPATH atlasbiblatex}
\usepackage{enumerate}
\usepackage{\ATLASLATEXPATH atlasphysics}
 
\graphicspath{{logos/}{figures/}}
 
\usepackage{ANA-TDAQ-2019-01-PAPER-defs}

\AtlasTitle{Performance of the upgraded PreProcessor of the ATLAS Level-1 Calorimeter Trigger}

\AtlasAbstract{
The PreProcessor of the ATLAS Level-1 Calorimeter Trigger prepares the analogue trigger signals sent from the ATLAS calorimeters by digitising, synchronising, and calibrating them to reconstruct transverse energy deposits, which are then used in further processing to identify event features.
During the first long shutdown of the LHC from 2013 to 2014, the central components of the PreProcessor, the Multichip Modules, were replaced by upgraded versions that feature modern ADC and FPGA technology to ensure optimal performance in the high pile-up environment of LHC Run~2.
This paper describes the features of the new Multichip Modules along with the improvements to the signal processing achieved.
}
 
\author{The ATLAS Collaboration}
 
\AtlasRefCode{TDAQ-2019-01}
 
\PreprintIdNumber{CERN-EP-2020-042}

\AtlasJournal{JINST}
\AtlasJournalRef{JINST 15 (2020) P11016}
\AtlasDOI{10.1088/1748-0221/15/11/p11016}

\AtlasCoverSupportingNote{Original Note and Discussions}{https://cds.cern.ch/record/2661780}
 
\AtlasCoverCommentsDeadline{1 April 2020}

\hypersetup{pdftitle={ATLAS document},pdfauthor={The ATLAS Collaboration}}
 
\begin{document}
 
\maketitle
 
\tableofcontents

\section{Introduction}
\label{sec:intro}
 
The Large Hadron Collider (LHC) at CERN accelerates protons to an energy of \SI{6.5}{\tera\electronvolt} and brings them into collision at four interaction points, with a frequency of \SI{40}{\mega\hertz}.
Installed at one of these points is the ATLAS detector \cite{PERF-2007-01, ATLAS-TDR-2010-19, PIX-2018-001}, which records the properties of particles produced in the collisions for later physics analysis.
The ATLAS experiment investigates a wide range of physics topics, with the primary focus of improving our understanding of the fundamental constituents of matter, and the interactions among them.
One of the most important goals for the first data-taking period of the LHC from 2009 to 2013 (Run~1) was to find and start measuring the properties of the long-sought Higgs boson, which was successfully discovered by the ATLAS and CMS collaborations in 2012~\cite{HIGG-2012-27, CMS-HIG-12-028}.
 
Due to the high rate of collisions, only a small subset of the produced events can be recorded for later analysis.
A powerful trigger system is thus required to select the events of interest.
As part of the ATLAS two-level trigger system, the  Level-1 Calorimeter (L1Calo) Trigger~\cite{l1calo_paper} identifies event signatures in the calorimeters.
The L1Calo PreProcessor (PPr) digitises the analogue trigger signals from the calorimeters, synchronises them, assigns them to the correct LHC bunch crossing (BC), determines their transverse energy ($E_{\textrm{T}}$) and transmits the results to two object-finding processor systems.
 
During LHC Run~1, the ATLAS trigger system including the L1Calo Trigger  operated efficiently at instantaneous luminosities of up to $8\times10^{33}$\,cm$^{-2}$\,s$^{-1}$ and primarily at centre-of-mass energies $\sqrt{s}= \SI{7}{\tera\electronvolt}$ and \SI{8}{\tera\electronvolt}~\cite{ppr_performance, PERF-2011-02, ATLAS-CONF-2012-048, TRIG-2012-01, PERF-2013-06, TRIG-2012-03}.
At the beginning of 2013, the LHC entered its first long shutdown (LS1) in order to increase the centre-of-mass energy to \SI{13}{\tera\electronvolt}, and the instantaneous luminosity by more than a factor two for its subsequent second data-taking period (Run~2, 2015--2018).
While this increased the discovery potential, for the detector it came with considerable challenges due to the increased rates of Standard Model processes.
Furthermore, the mean number of interactions per bunch crossing, which during Run~1 was on average 21 and peaked at around 40, increased to an average of 38, and at times exceeded 60~\cite{DAPR-2013-01, ATLAS:2019pzw}.
 
During LS1, the L1Calo Trigger was upgraded to account for the intense conditions expected for Run~2, and to continue operating at the high efficiency achieved during Run~1.
For the PPr, the ASIC-based Multichip Modules (MCMs), which perform the main signal processing tasks, were replaced by FPGA-based new MCMs (nMCMs).
These new modules improve on the design of the Run~1 hardware and allow for the implementation of enhanced signal processing algorithms \cite{tdaq_tdr}.
 
This paper describes the nMCM as operated in Run~2, focussing specifically on the improved performance achieved relative to the Run~1 PPr.
The overall performance of the ATLAS trigger system during Run~2 is highlighted in~\cite{TRIG-2016-01, egammatrig, Aad:2020les, Aad:2020uyd}.
An overview of the ATLAS detector and trigger system is given in Section~\ref{sec:atlas}.
Section~\ref{sec:ppr_run2} introduces the L1Calo Trigger.
The hardware and firmware implementation of the nMCMs is summarised in Section~\ref{sec:nmcm}, and the performance improvements are discussed in Section~\ref{sec:performance}.
A summary is provided in Section~\ref{sec:summary}.

\section{ATLAS detector and trigger system}
\label{sec:atlas}
ATLAS is a multipurpose particle detector utilising a forward--backward symmetric cylindrical geometry to achieve near 4$\pi$ coverage in solid angle.\footnote{
ATLAS uses a right-handed coordinate system with its origin at the nominal interaction point in the centre of the detector and the $z$-axis along the beam pipe. The $x$-axis points from the IP to the centre of the LHC ring, and the $y$-axis points upwards. Cylindrical coordinates ($r$, $\phi$) are used in the transverse plane, $\phi$ being the azimuthal angle around the $z$-axis. The pseudorapidity is defined in terms of the polar angle $\theta$ as $\eta = - \ln \tan(\theta/2)$.}
The three major subcomponents of ATLAS are an inner tracking detector, the surrounding electromagnetic and hadronic calorimeters, and a muon spectrometer.
 
The inner tracking detector covers the pseudorapidity range $|\eta| < 2.5$.
Consisting of silicon pixel, silicon microstrip and transition radiation tracking detectors, the inner tracking detector allows the trajectories of charged particles produced in the proton--proton ($pp$) interactions to be reconstructed with high efficiency.
It is embedded in a superconducting solenoid providing a \SI{2}{\tesla} axial magnetic field.
 
The electromagnetic (EM) calorimeter is a lead/liquid argon (LAr) sampling calorimeter with an accordion geometry.
It is divided into a barrel section (EMB) covering the pseudorapidity region $|\eta| < 1.475$
and two endcap sections (EMEC) covering $1.375 < |\eta| < 3.2$, immersed in three LAr-filled cryostats.
The EMB itself is split into two half-barrel modules which cover the positive and negative $\eta$ regions.
The barrel and endcap sections also feature a thin LAr presampler layer (PS), covering the pseudorapidity interval $|\eta| < 1.8$,
which is used to correct for energy loss upstream of the calorimeter.
The hadronic (HAD) calorimeter, surrounding the EM calorimeter, consists of an iron/scintillator tile (Tile) calorimeter
in the range $|\eta| < 1.7$ and two copper/LAr endcap (HEC) calorimeters spanning $1.5 < |\eta| < 3.2$.
The calorimeter acceptance is extended by two copper/LAr and tungsten/LAr forward calorimeters (FCALs) ranging up to $|\eta| = 4.9$,
and hosted in the same cryostats as the EMEC and HEC.
 
The ATLAS calorimeters have a highly granular lateral and longitudinal segmentation.
Including the PS, the EM calorimeter is segmented into four layers longitudinal in shower depth (PS, three layers in EMB, three in EMEC).
The first active layer after the PS is finely segmented in the $\eta$ direction, offering excellent discrimination between isolated photons and closely spaced photons from the decays of neutral hadrons.
For electrons and photons with high transverse energy, most of the energy is collected in the second layer, which has a granularity of $0.025 \times 0.025$ in $\Delta \eta \times \Delta \phi$.
The third layer provides energy measurements in the tails of high-energy EM showers.
Within the HAD calorimeter, Tile is segmented into three longitudinal layers, while the HEC calorimeters consist of four layers.
The granularity in $\Delta \eta \times \Delta \phi$ ranges from typically $0.1 \times 0.1$ in the region $|\eta| < 2.5$ to $0.2 \times 0.2$ for the interval $2.5 < |\eta| < 3.2$.
In the forward regions, each FCAL consists of three modules. The first module (FCAL1) is optimised for EM measurements,
while the two latter modules (FCAL2 and FCAL3) provide measurements for HAD interactions.
 
The muon spectrometer is located outside of the calorimeter systems and is based on three large air-core toroid superconducting magnets with eight coils each.
It includes precision tracking chambers providing accurate muon tracking for $|\eta| < 2.7$ and fast-triggering detectors covering $|\eta| < 2.4$.
 
ATLAS uses a two-level trigger system to select events for permanent storage~\cite{TRIG-2016-01}.
It consists of a hardware-based first-level (L1) trigger and a software-based High-Level Trigger (HLT).
The L1 trigger uses custom electronics that analyses a subset of the detector information for each BC at reduced granularity.
From the raw \SI{40}{\mega\hertz} bunch crossing rate, it accepts potentially interesting events up to a design maximum rate of \SI{100}{\kilo\hertz}.
This has to be accomplished within a total latency of \SI{2.5}{\micro\second} after the collision occurred, including signal transmission to the off-detector trigger processor system and the return of the accept/reject decision to on-detector electronics.
 
Within the L1 system, the L1 Calorimeter and L1 Muon Triggers process data from the respective sub-detectors to identify candidates for high-$E_{\textrm{T}}$ particles.
Based on the properties of these candidates, the L1 Topological (L1Topo) trigger analyses the topology of events.
The results of the different L1 trigger systems are collected by the Central Trigger Processor (CTP), which forms the final L1 decision, and issues an L1 Accept (L1A) signal for selected events.
L1 also supplies information to the HLT in the form of so-called Regions-of-Interest (RoIs) which contain the candidates, including their geometrical $\eta$--$\phi$ locations within the detector and the criteria (e.g.\ thresholds) that they satisfied.
 
Upon reception of the L1A signal, the Data Acquisition (DAQ) system collects the data for the selected event from the ATLAS sub-detectors and buffers them.
The buffered data are read out either regionally for the identified RoIs only, or globally for the full detector, and sent to the nodes of the HLT farm for processing.
The HLT uses fine-granularity calorimeter information, precision measurements from the muon spectrometer and tracking information from the inner detector, in order to refine the selections made by L1.
HLT reconstruction can be executed either within the RoIs identified at L1 or for the full detector.
Events passing the HLT are permanently stored for offline analysis, for Run~2 with an average rate of \SI{1}{\kilo\hertz} and event sizes of approximately \SI{1.5}{\mega\byte}.

\section{Level-1 Calorimeter Trigger}
\label{sec:ppr_run2}
 
\begin{figure}[t]
\centering
\includegraphics[width=0.7\textwidth]{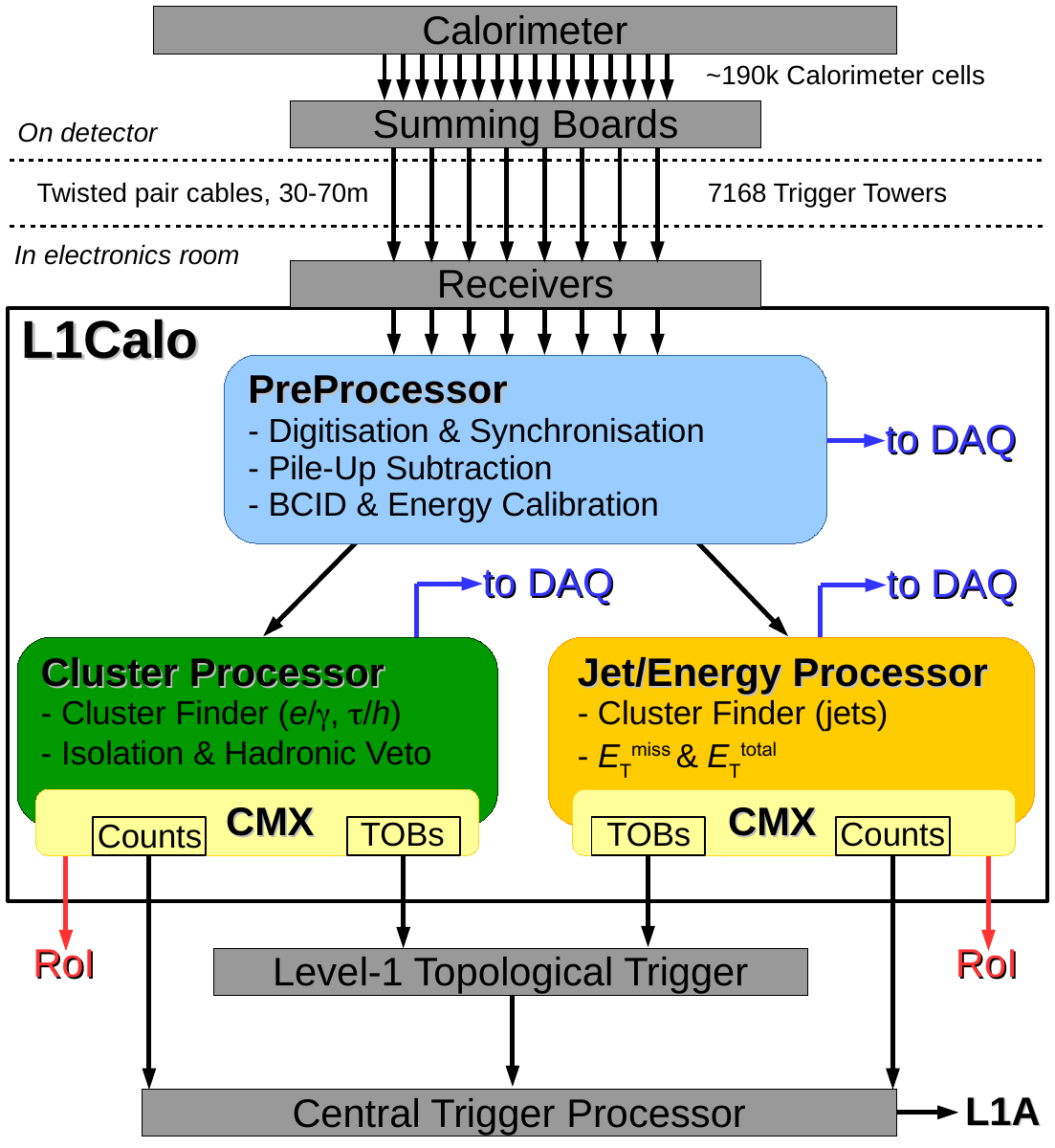}
\caption[L1Calo Run~2 Architecture]{Architecture of L1Calo Trigger in Run~2.}
\label{fig:L1CaloPhase1}
\end{figure}
 
The L1Calo Trigger identifies candidates for high-$E_{\textrm{T}}$ particles in the ATLAS calorimeters and computes global energy sums for the full event, such as the missing transverse momentum ($E_{\textrm{T}}^{\textrm{miss}}$) and total $E_{\textrm{T}}$ ($E_{\textrm{T}}^{\textrm{total}}$).\footnote{
Here, $E_{\textrm{T}}^{\textrm{miss}}$ is the negative of the vectorial sum, and $E_{\textrm{T}}^{\textrm{total}}$ is the scalar sum, of all
energy depositions in the transverse plane.}
The L1Calo electronics is located entirely off the detector in the  separate ATLAS electronics room, and data processing corresponds to a fixed latency of about \SI{1}{\micro\second}.
Therefore, the largest fraction of the total L1 time budget of \SI{2.5}{\micro\second} is consumed by the signal transmission delays from and back to the detector front-end electronics.
 
Figure~\ref{fig:L1CaloPhase1} shows a schematic overview of the L1Calo Trigger as operated during Run~2.
It comprises a preprocessing stage and two object-finding processors.
Its input data comes from 7168 analogue signals at reduced granularity, called Trigger Towers (TTs),
from both the ATLAS EM and HAD calorimeters (see Section~\ref{sec:inputpath}).
 
The PPr digitises the analogue TT pulses from the calorimeters, identifies those which correspond to significant energy depositions,
and applies the final $E_{\textrm{T}}$ calibration and signal preparation for use in the subsequent algorithmic processors.
The Cluster Processor (CP) searches for electron, photon and $\tau$-lepton candidates in the form of narrow clusters within the calorimeters.
Isolation criteria and vetoes on energy deposits in the hadronic calorimeter layer (hadronic vetoes) are also used to distinguish EM from HAD showers.
The Jet-Energy Processor (JEP) finds and determines the properties of jet candidates and computes the global energy sums.
Both processors use sliding-window algorithms in the $\eta$--$\phi$ space of the TTs to find local maxima of energy deposits.
Details of the CP and JEP systems are given in Ref.~\cite{l1calo_paper}.
 
The particle candidates identified by the processors are collected in Common Merger Modules (CMXs), which transmit them to the L1Topo trigger system in the form of Trigger Objects (TOBs).
The CMXs also apply $E_{\textrm{T}}$ thresholds as defined in a programmable trigger menu to the TOBs, and transmit the resulting multiplicity of particle candidates above thresholds to the CTP.
 
Upon reception of an L1A signal from the CTP, the L1Calo system provides information to the DAQ system via dedicated Readout Drivers (RODs).
This comprises both the RoIs of the identified particle candidates used to seed the HLT algorithms, and results from intermediary processing steps to be stored for offline analysis.
The latter are essential for calibration and performance studies and allow for the offline reconstruction of the online L1Calo results, which is used for error monitoring.
 
\subsection{Input signal path}
\label{sec:inputpath}
 
The two ATLAS EM and HAD calorimeters consist of approximately 190k cells in total, which are combined into 7168 projective TTs by grouping calorimeter cells in defined $\eta$--$\phi$ regions~\cite{lar_tdr, tile_tdr, Buchanan_2008, CERQUEIRA2007117}.
TTs are formed independently for the EM and HAD calorimeter layers, with the granularity being nominally $\Delta \eta \times \Delta \phi = 0.1 \times 0.1$ for the barrels and outer endcaps ($|\eta|<2.5$), mostly $0.2 \times 0.2$ in the inner endcaps ($2.5<|\eta|<3.2$) and $0.4 \times 0.4$ in the FCALs.
The analogue signals of all cells within a TT are summed within the calorimeter front-end, and the resulting signals are transferred via 30 to \SI{70}{\meter} long twisted-pair cables to the ATLAS electronics room.
 
In the electronics room, the TT signals are routed to Receiver modules.
These compensate for signal attenuation and provide variable-gain amplifiers to scale the signals, e.g.\ to convert them from energy $E$ to $E_{\textrm{T}}$ for the HAD calorimeters, while for the EM calorimeters this is performed in the front-end electronics.
The amplifiers are also used to calibrate all signals to a common energy scale, given that the responses from the different calorimeter materials to different types of particles are not equal (see Section~\ref{sec:preprocessor}).
The scaled analogue pulses are then transmitted to the L1Calo PreProcessor.
 
\begin{figure}[t]
\centering
\includegraphics[width=0.65\textwidth]{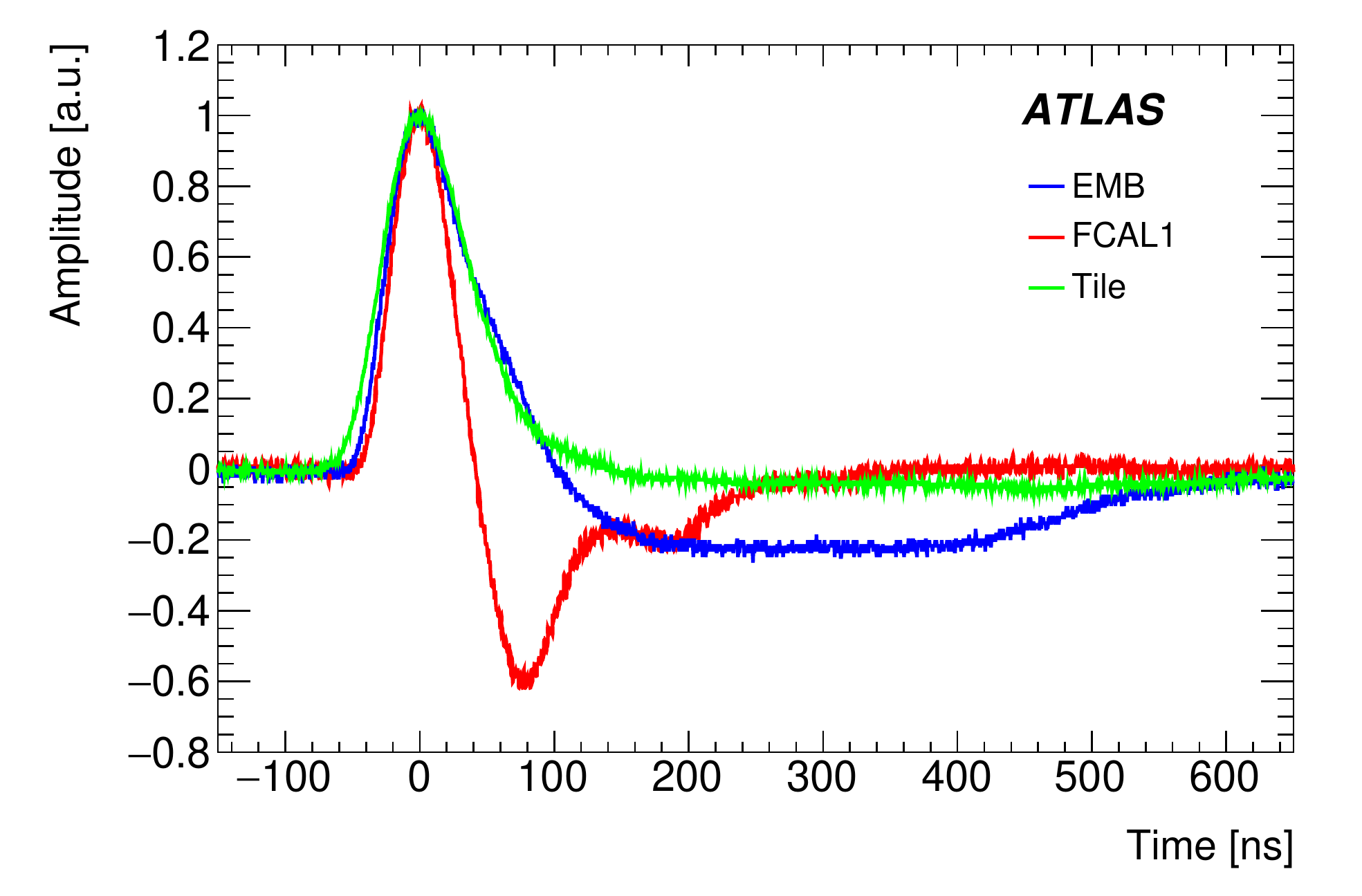}
\caption[TT Pulse Shapes]{Normalised pulse shapes for TTs from the EM barrel and FCAL as well as the HAD barrel calorimeter regions, recorded with an oscilloscope during $pp$ collisions at the Receiver output.}
\label{fig:Pulses}
\end{figure}
 
Figure~\ref{fig:Pulses} shows examples of TT signal shapes from different calorimeter regions.
The LAr signals, which account for about 75\% of the signals received by the PPr, are bipolar with a large positive peak followed by a long negative undershoot.
The signals from the Tile calorimeter have a unipolar shape.
The pulses span multiple BCs, which are spaced every \SI{25}{\nano\second}.
While the exact shape varies strongly in $\eta$ and between the EM and HAD layers of the calorimeter, its variation in $\phi$ is usually small.
As result, TTs with the same $\eta$ coordinate in the same calorimeter layer are in general treated equivalently in the calibration of the PPr.
 
Because the size of the TTs varies in $\eta$, L1Calo configuration parameters and calibration results are often organised in linear $\eta$-bins.
Each bin contains all TTs with the same $\eta$-coordinate, with the most central $\eta$-coordinate being defined as bin number 0.
In total, there are 66 integer $\eta$-bins, ranging from $-33$ to $+33$.
 
As a consequence of the high instantaneous luminosities achieved by the LHC, multiple $pp$ interactions occur within each BC (pile-up).
The products of these additional interactions typically deposit small amounts of energy and contribute to the formation of an omnipresent background to the signals from the hard-scattering processes.
Two types of pile-up, based on their origin in time, affect the detector performance.
In-time pile-up refers to effects which arise in the same BC as the hard interaction.
Out-of-time pile-up refers to effects due to contributions from signals in neighbouring BCs as a result of the analogue pulses spanning over multiple BCs.
Therefore, out-of-time pile-up effects depend strongly on the pulse shape and the detector regions.
 
\subsection{PreProcessor system}
\label{sec:preprocessor}
 
The PPr is the initial stage of the L1Calo system to receive the TT signals.
It consists of 124 functionally identical PPr Modules (PPMs), organised in 8 VME crates in the ATLAS electronics room.
The details of the PPr system can be found in Ref.~\cite{ppr_performance}.
 
\begin{figure}[t]
\centering
\includegraphics[width=0.8\textwidth]{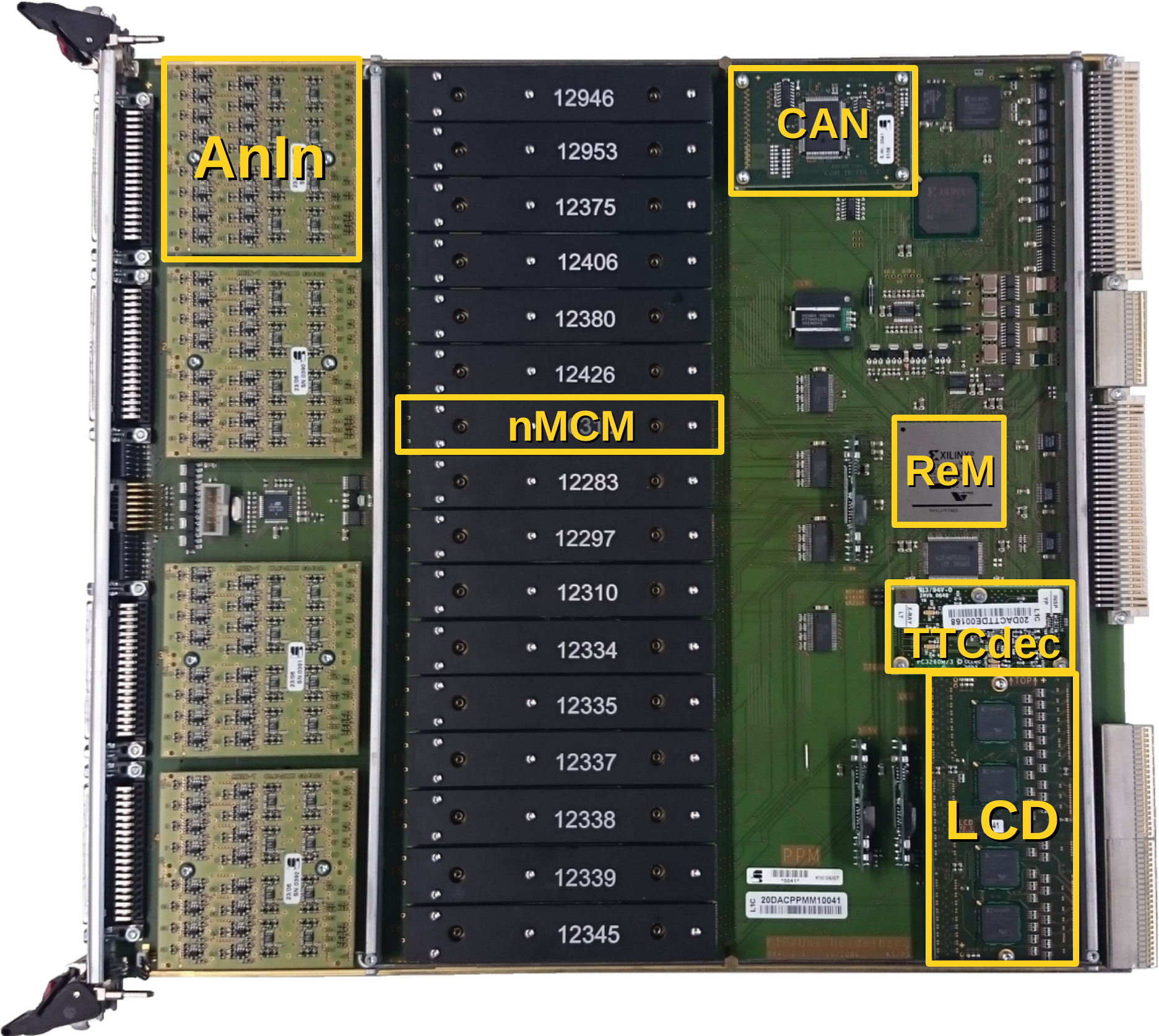}
\caption[PreProcessor Module (PPM)]{Photograph of a PPM, highlighting the different types of components as described in Section~\ref{sec:preprocessor}.
This PPM is already equipped with new Multichip Modules, which are covered by heat sinks, labelled with the unique serial number of the chip (see Section~\ref{sec:nmcm_hw}). }
\label{fig:PpmPic}
\end{figure}

Figure~\ref{fig:PpmPic} shows a photograph of a PPM.
Each PPM receives 64 TT signals, which are transmitted through the four input connectors on the left side.
For each connector, one Analogue Input (AnIn) board is installed, which converts the terminated differential input signals to single-ended signals, and scales them such that they fit the dynamic range of the downstream devices.
In addition, it applies DC offsets to equalise the baseline among the different inputs.
 
The AnIn output signals are routed to 16 MCMs.
The MCMs digitise the TT signals, with the gain factors set in the Receiver system such that one ADC count corresponds to $E_{\textrm{T}}$ of \SI{250}{\mega\electronvolt}.
The digital signals are synchronised and assigned to the BC corresponding to the one containing the signal peak.
 
The MCMs send digital $E_{\textrm{T}}$ results via two separate serial high-speed data streams to a Low-Voltage-Differential-Signalling (LVDS) Cable Driver (LCD).
The LCD performs the fan-out and pre-compensation of the signals required for LVDS transmission at \SI{480}{\mega\bit/\second} to the CP and JEP modules,
which can be as far as \SI{15}{\meter} away.
 
The control and configuration of the PPM is handled by the Readout Merger (ReM) FPGA, which is accessible through the VME interface.
For each L1A, the ReM FPGA also collects data related to the triggered event from all MCMs and transmits them to the ROD, via a Rear GLink Transition Module that is installed in the rear part of the VME crate and connected to the PPM via a backplane connector.
 
The TTC Decoder (TTCdec) card receives the LHC clock and the L1A signal from the Timing and Trigger Control (TTC) system, and forwards them to the ReM FPGA for distribution to the remaining PPM components.
The Controller Area Network (CAN) card collects temperature and voltage information from the other onboard devices and supplies them to the ATLAS Detector Control System via CAN bus.

\section{New Multichip Module}
\label{sec:nmcm}
 
During LS1, the original MCMs were replaced by new MCMs (nMCMs).
The nMCMs are electrically and mechanically equivalent replacements for the MCMs.
Their design is optimised using the experience gained from Run~1 operation, with modern components allowing for an extended firmware implementation to improve the PPr performance.
The following subsections give an overview of both the hardware and firmware implementation of the nMCM.

\subsection{Hardware implementation}
\label{sec:nmcm_hw}
 
\begin{figure}[p!]
\centering
\centering
\includegraphics[width=0.85\textwidth]{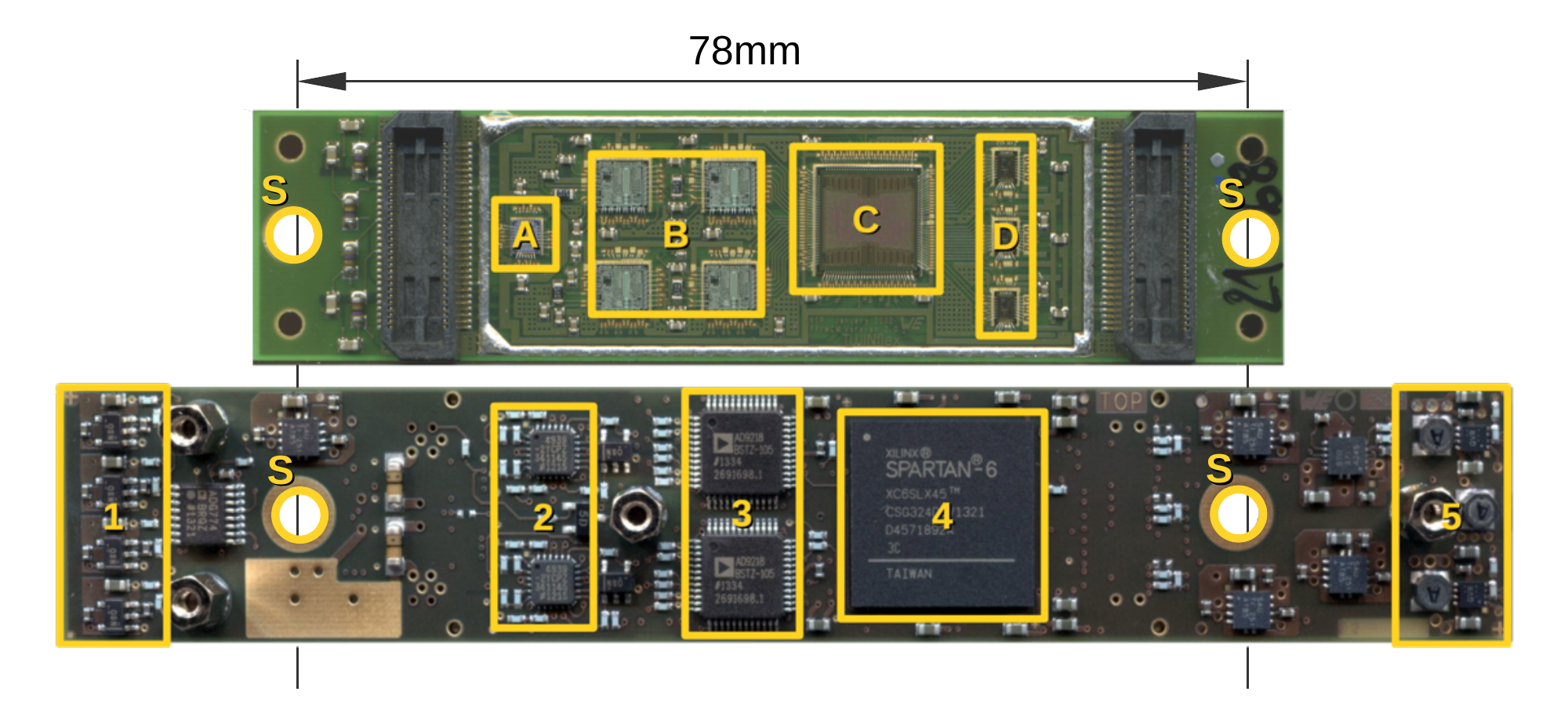}
\caption[New Multichip Module (nMCM)]{Photographs of the original (top) and new (bottom) Multichip Module. The highlighted components are: (A) the PHOS4 chip, (B) the FADCs, (C) the PPrASIC and (D) the LVDS transmitters on the MCM; (1) the signal generator circuit, (2) the operational amplifiers, (3) the FADCs, (4) the CALIPPR FPGA (Xilinx Spartan-6) and (5) the power converters on the nMCM. The EEPROM device is mounted on the far side of the nMCM. (S) indicate the screw holes for mounting on the PPM. The nMCM is slightly longer than the MCM.}
\label{fig:nmcmPic}
\end{figure}

Figure~\ref{fig:nmcmPic} shows a comparison between the MCM and the nMCM.
On the MCM, four Flash ADCs (FADCs) digitise the TT signals at \SI{40}{\mega\hertz} with a PHOS4 chip \cite{phos4_paper} to
fine-tune the sampling delays in steps of \SI{1}{\nano\second}. 
The digital signals are then processed within the PPrASIC \cite{asic_manual}, which performs the main PPr functions described in Section~\ref{sec:preprocessor}.
The resulting $E_{\textrm{T}}$ values are transmitted off the MCM by three LVDS serialisers.
 
\begin{figure}[p!]
\centering
\includegraphics[width=0.85\textwidth]{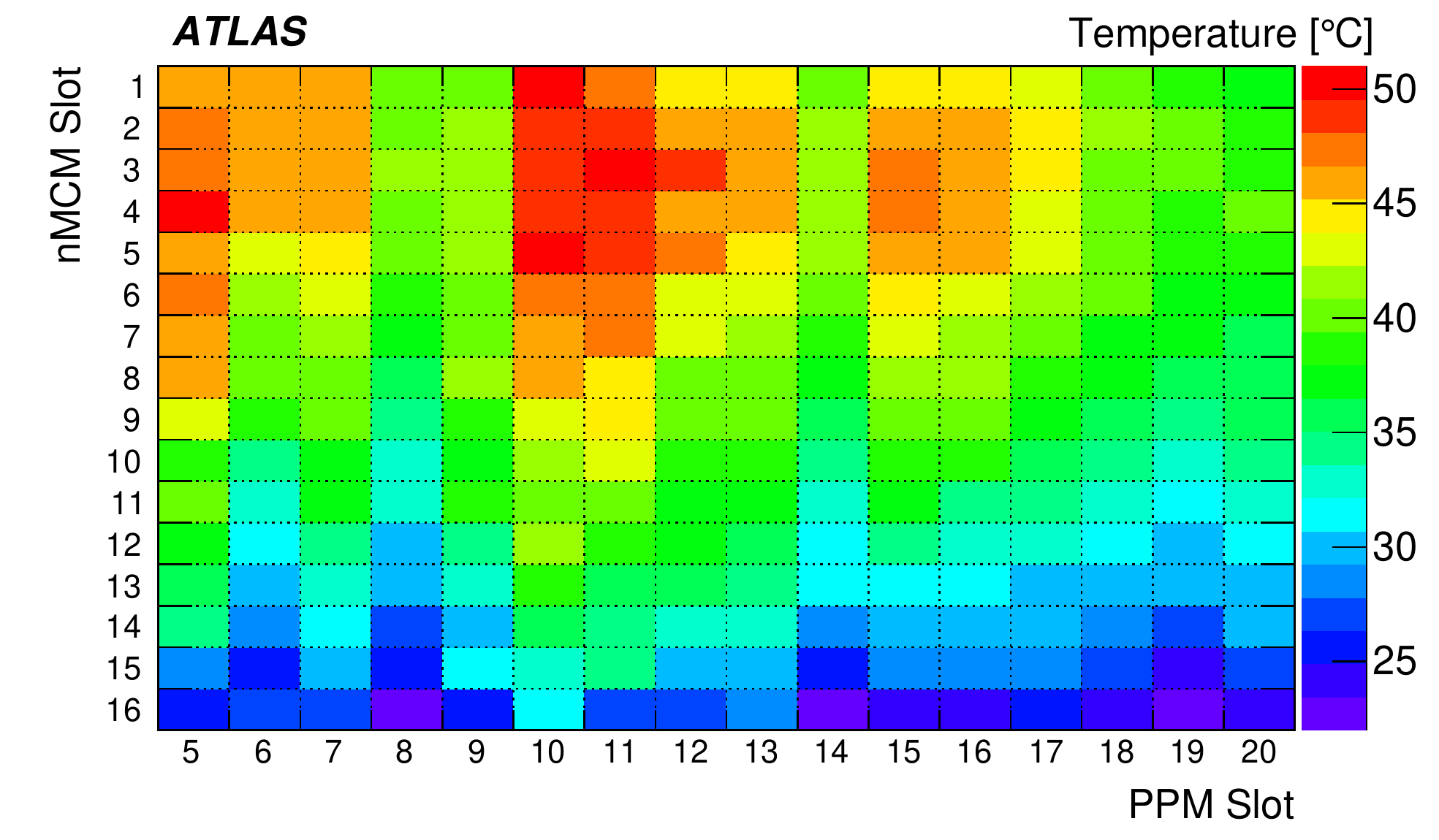}
\caption[MCM Temperatures]{Temperatures of all nMCMs in a PPr crate. This measurement was performed in a test crate, operating all nMCMs in playback mode (cf.\ Section~\ref{sec:nmcm_fw}) to run stress patterns. Each rectangle represents one nMCM, each column a fully loaded PPM.
Circular fans at the bottom of the crate force a high-flux air stream upwards and are centred around slots 6 and 16, covering slots 3--9 and 13--19, respectively.}
\label{fig:nmcmTemp}
\end{figure}
 
On the nMCM, two dual-channel FADCs digitise the analogue input signals at \SI{80}{\mega\hertz}. 
Two dual-channel operational amplifiers adjust the nMCM input signals so that they fit the slightly different voltage requirements of the new FADCs, including conversion from single-ended to differential signalling. 
The digitised signals are routed to the Calorimeter Information PreProcessor (CALIPPR) FPGA.
It combines and improves upon the functions provided on the MCM by the PPrASIC, the PHOS4 chip and the LVDS transmitters, also adding new features in the signal processing chain (see Section~\ref{sec:nmcm_fw}).
Furthermore, a set of power converters is installed to adapt the supply voltages for the different requirements of the nMCM compared to the MCM components.
An on-board signal generation circuit allows the nMCM functionality to be tested independently of external inputs.
 
The CALIPPR firmware bit file is stored in an EEPROM and downloaded to the FPGA automatically when powering on the modules.
In case of updates, a new bit file can be written to the EEPROM from VME.
In addition, the EEPROM stores a unique, write-protected identification number for each nMCM.
 
The nMCM has a total power consumption of \SI{3.7}{\watt}, which is dissipated via an anodised aluminium heat sink.
The heat sink covers the whole nMCM (cf.\ Figure~\ref{fig:PpmPic}), and is connected to the active components in the centre of the nMCM by a heat conducting paste, leading to optimal cooling for these hottest components on the module.
 
Figure~\ref{fig:nmcmTemp} shows the temperatures of all nMCMs in a PPr crate.
The nMCMs reach stable operating temperatures between \SI{20}{\celsius} and \SI{55}{\celsius}, depending on their position relative to the crate-cooling fans, which are located along the bottom of the crate and produce a vertical upward air stream.
As a result, the hottest modules are positioned at the top of the crate and in slots that are situated far from the centre of the cooling fans along the horizontal axis.
Despite the variation in temperature, neither performance nor lifetime differences have been observed for the hardware components of modules in the different slots.

\subsection{Firmware overview}
\label{sec:nmcm_fw}
 
\begin{figure}[t]
\centering
\includegraphics[width=\textwidth]{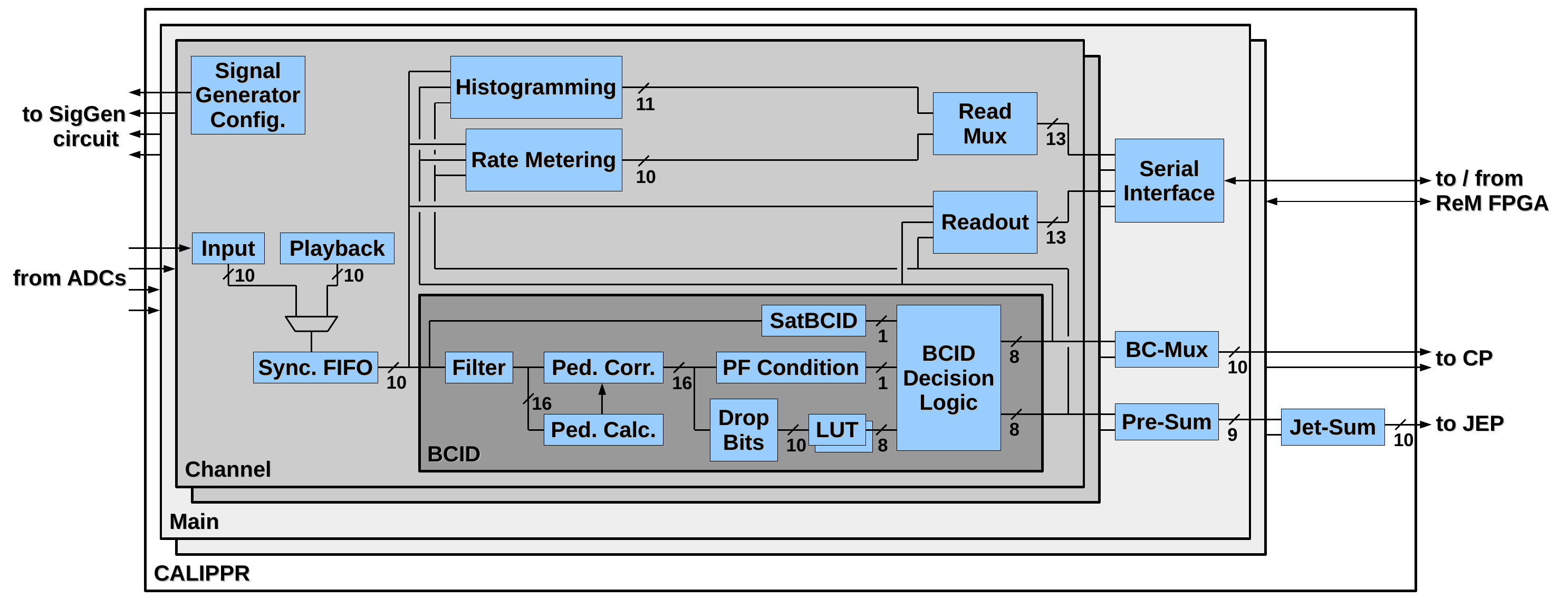}
\caption[CALIPPR Firmware Overview]{Overview of the CALIPPR firmware design.}
\label{fig:calippr}
\end{figure}

The CALIPPR firmware implementation is based on the design of the PPrASIC, which is described in Refs.~\cite{asic_manual, ppr_performance}.
Figure~\ref{fig:calippr} shows a schematic overview of the main processing elements.
Details of the functionality and performance are given in Section~\ref{sec:performance}.
 
The design is organised into four identical `Channel' logic blocks, each processing in parallel the signal from one of the four input TTs.
The FADC output is received in the `Input' block, which latches it with the internal clock.
Here also, the \SI{80}{\mega\hertz} input is reduced to a \SI{40}{\mega\hertz} stream by dropping every other ADC value, as required by most of the processing algorithms.
Only the `SatBCID' (Saturated Bunch Crossing Identification) and `Readout' logic blocks use the full \SI{80}{\mega\hertz} data.
The subsequent `Sync.\ FIFO' block applies a configurable delay that is used to synchronise the different TT signals across the whole PPr system.
 
The synchronised signals are then processed by the `BCID' (Bunch Crossing Identification) logic block.
The main BCID algorithm is a Peak-Finder, which is optimised for low-$E_{\textrm{T}}$ signals.
It is based on a Finite Impulse Response (FIR) filter, which as a first step improves the signal-to-noise ratio in the `Filter' block.
A new Pedestal Correction algorithm, applied in the subsequent `Ped.\ Corr.' block, corrects the filter output stream for pile-up effects before it is searched for peaks in the `PF Condition' block.
A positive result identifies the corresponding BC as the origin of the collision of interest.
 
The corrected filter output is calibrated to $E_{\textrm{T}}$ in a configurable Look-Up-Table (LUT) memory.
In the LUT a noise threshold is applied to reduce the impact of electronic noise and effects of pile-up.
The CALIPPR design introduces a second `LUT', allowing for independent $E_{\textrm{T}}$ calibrations for the CP and the JEP.
 
TT signals that correspond to $E_{\textrm{T}}$ larger than approximately \SI{250}{\giga\electronvolt} saturate the FADCs on the nMCM.
In this case, the `LUT' output is set to its maximum value.
A dedicated `SatBCID' algorithm is used to assign these signals to the correct BC, making use of the higher digitisation frequency of the FADCs.
 
The results of the two BCID algorithms are combined in the `BCID Decision Logic' block, forming the final BCID decision for each BC.
If this decision is positive, the `LUT' output is forwarded to the CALIPPR output formatting blocks, otherwise it is set to 0.
 
The $E_{\textrm{T}}$ results of the `BCID' logic are formatted and then transmitted to the CP and JEP systems.
For the CP, two TT results are multiplexed on the same link in two subsequent BCs in the `BC-Mux' block \cite{l1calo_paper}.
For the JEP, the $E_{\textrm{T}}$ results of the four nMCM channels are summed and the result is transmitted on a single link.
 
The functionality and performance of the nMCM are constantly monitored using several different methods.
Results from the different processing stages are stored in `Readout' pipeline memories and read out into the ATLAS DAQ system for each triggered BC.
The CALIPPR implementation adds information concerning the new functionality to the readout stream.
This includes the Pedestal Correction, the second LUT and decision bits of the improved SatBCID algorithm, which allow the commissioning and monitoring of this algorithm to be performed.
Also, the possibility to read out the \SI{80}{\mega\hertz} ADC data stream is provided.
 
Dedicated blocks in the firmware provide rate measurements and histograms of the $E_{\textrm{T}}$ results that are not biased by L1 event selection criteria.
These `Rate Metering' and `Histogramming' functions have been crucial to the operation of the PPr during both Run~1 and Run~2, providing quick diagnostics of faulty channels.
 
A `Playback' memory allows the digital algorithms to be tested by injecting known signals into the processing chain.
An additional on-board test facility introduced on the nMCM is the signal generator, which allows for similar checks of the analogue components.
 
Each group of two channels shares a bidirectional serial interface to the ReM FPGA, and is organised in a separate `Main' block.
The interface is used to transmit configuration data from the ReM FPGA to the CALIPPR FPGA, and to retrieve the data collected in the readout and monitoring blocks in the other direction.

\section{Improved signal processing on the nMCM}
\label{sec:performance}
 
The major motivation for the changes in both hardware and firmware design between the MCM and the nMCM is a performance improvement of the PPr system.
This Section discusses those aspects of the signal processing that have been addressed by the upgrade.

\subsection{Electronic noise}
\label{sec:performance_noise}
 
During the operation of the MCM in Run~1, as well as in the development phase of the nMCM, several sources of electronic noise were identified and removed with the introduction of the new module.
Those arose both from characteristics of the MCM electronics itself and from its implementation on the PPM motherboard.
In particular, the nMCM allows for the minimisation of effects originating from the coupling between analogue and digital ground planes on the PPM by reducing the number of digital signals required by the MCM:
 
\begin{itemize}
\item[(a)] Only a single clock signal is transmitted from the ReM FPGA to the CALIPPR FPGA. The remaining clock signals are generated in PLLs (phase-locked loops) included in the CALIPPR design.
\item[(b)] The protocol of the serial interface between the ReM FPGA and the CALIPPR FPGA was reworked such that its idle state word is a constant zero, instead of containing a single active bit.
\end{itemize}
 
\begin{figure}[t]
\centering
\begin{subfigure}[b]{0.49\textwidth}
\centering
\includegraphics[width=\textwidth]{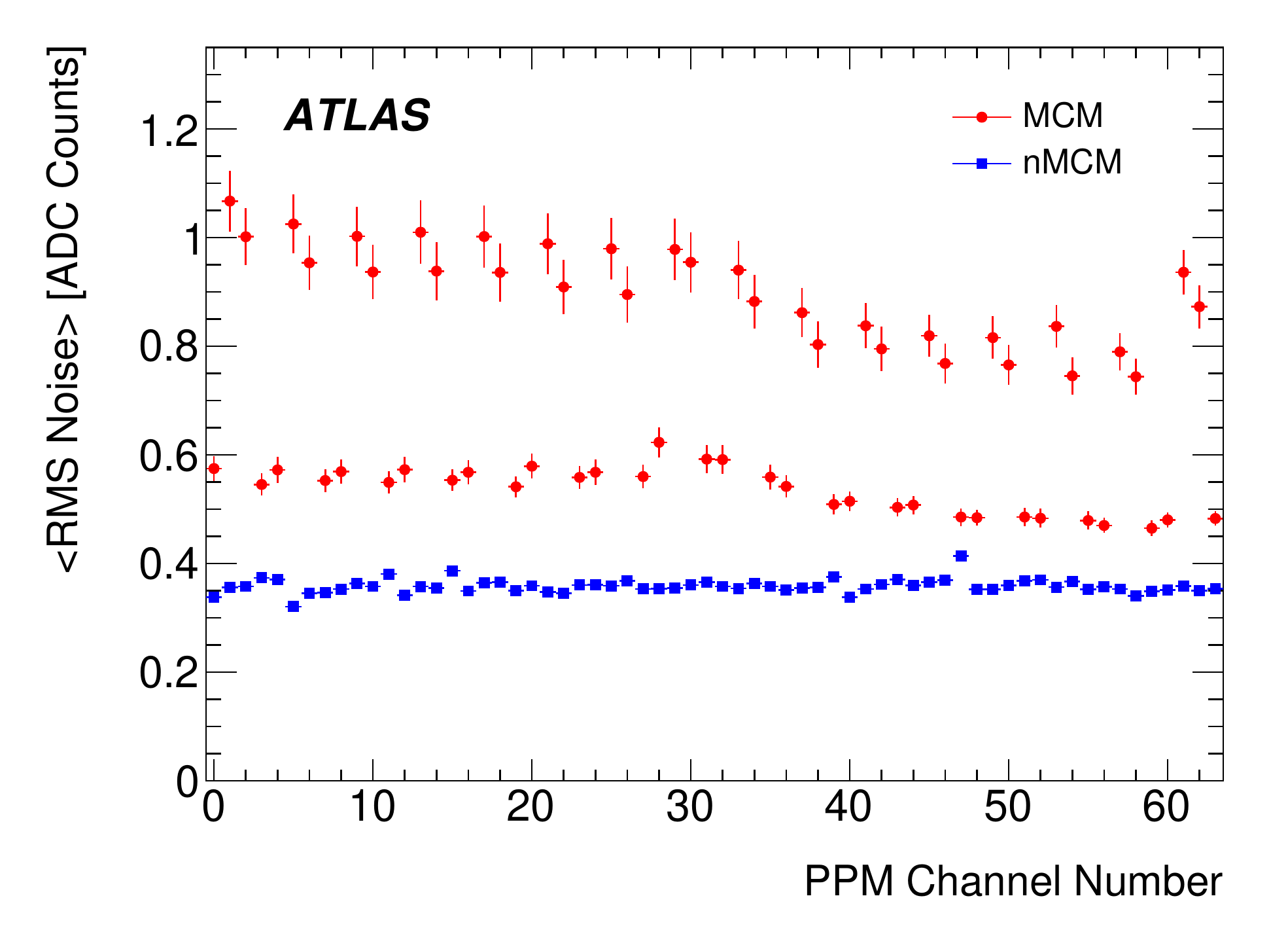}
\caption[]{}
\end{subfigure}
~
\begin{subfigure}[b]{0.49\textwidth}
\centering
\includegraphics[width=\textwidth]{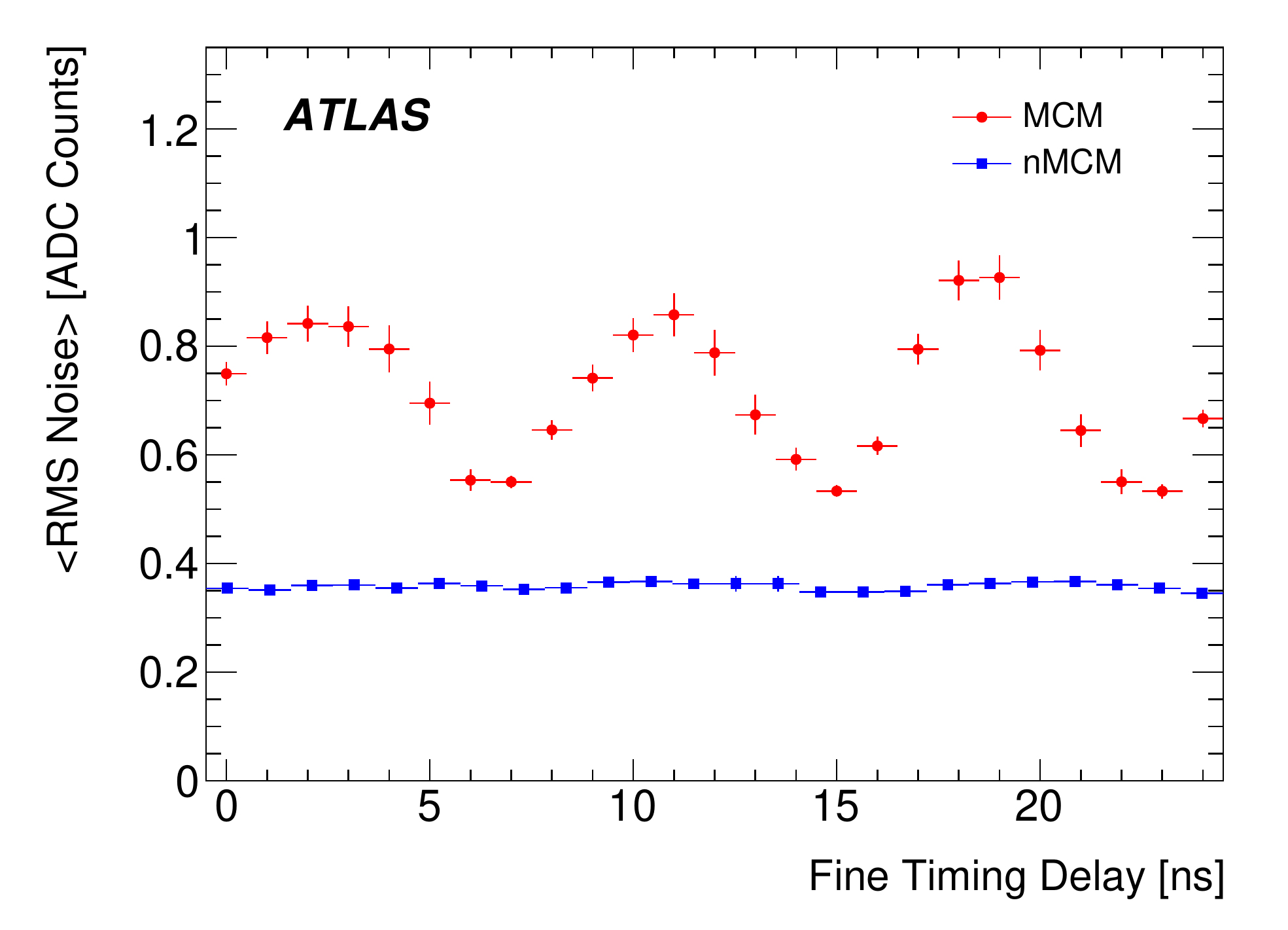}
\caption[]{}
\end{subfigure}
\caption[Noise on PPM]{(a) The average noise for all channels in a PPM with no input cables connected, for the two cases of a PPM fully equipped with MCMs or nMCMs. For the MCM, two of the four channels on each module are noisier, leading to two separate curves. (b) The noise averaged over all channels in a PPM measured with different fine timing settings, for both the MCM and the nMCM. It should be noted that the nMCM implementation uses one less fine-timing step than the MCM (see Section~\ref{sec:performance_timing}).}
\label{fig:NoisePpm}
\end{figure}
 
Figure~\ref{fig:NoisePpm}(a) shows the RMS (root mean square) of the ADC output distribution for all channels of one PPM, averaged over multiple measurements.
The measurement was performed in a test crate with no input cables connected, for both a PPM equipped with MCMs and a PPM with nMCMs.
The comparison demonstrates the noise reduction achieved with the nMCM.
 
\begin{figure}[p!]
\centering
\begin{subfigure}[b]{0.49\textwidth}
\centering
\includegraphics[width=\textwidth]{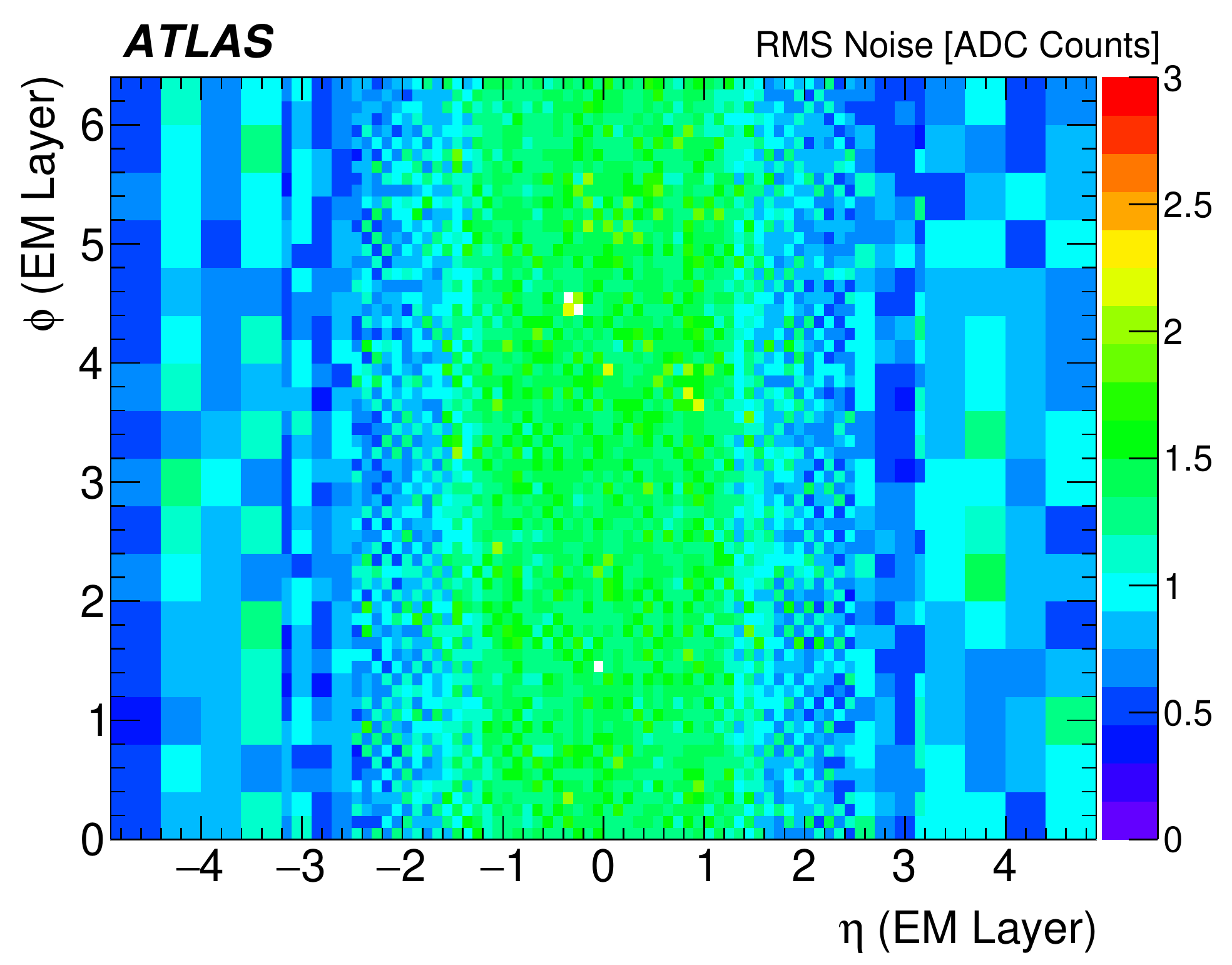}
\caption[]{}
\end{subfigure}
~
\begin{subfigure}[b]{0.49\textwidth}
\centering
\includegraphics[width=\textwidth]{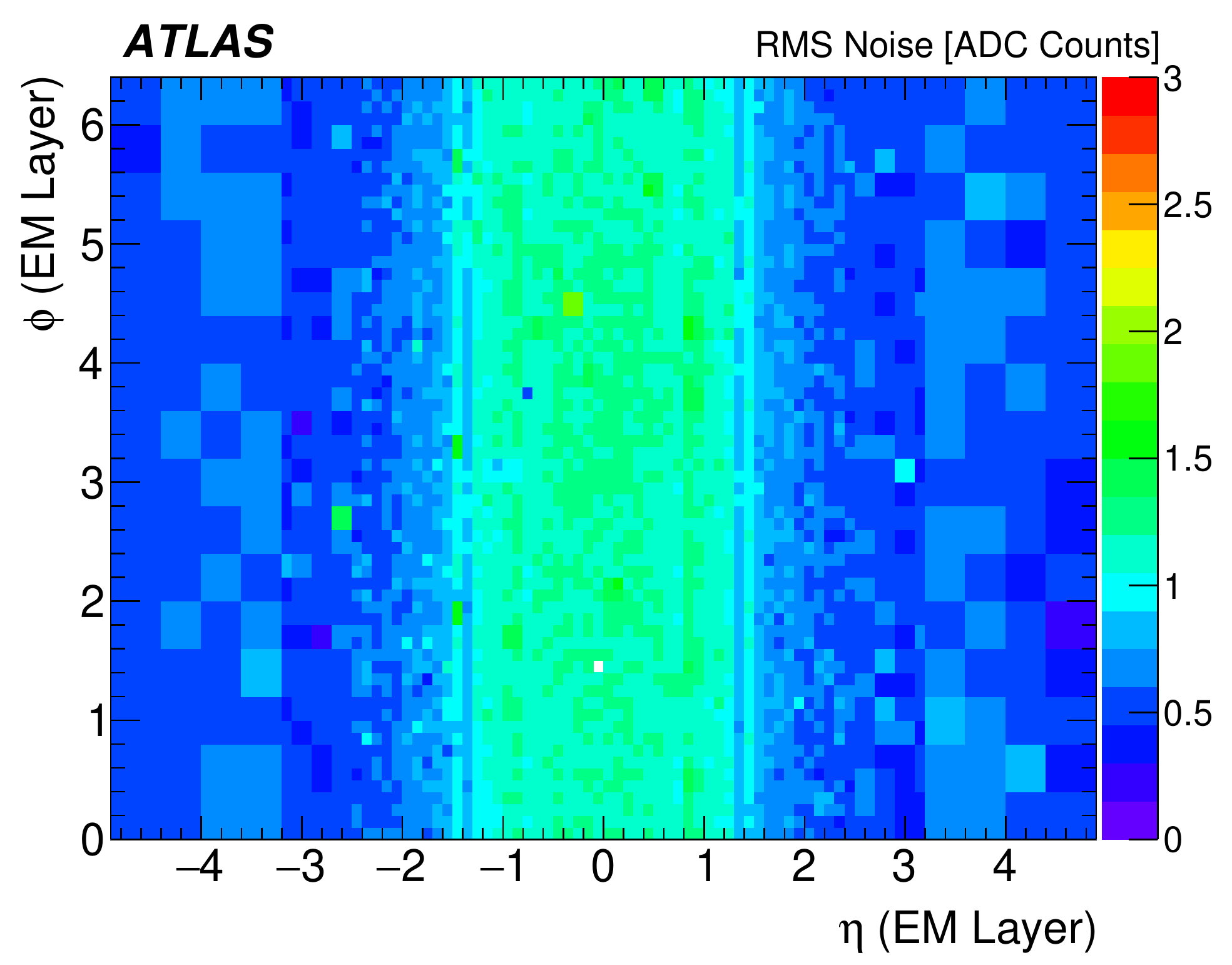}
\caption[]{}
\end{subfigure}
\caption[Total Noise]{$\eta$--$\phi$-maps of the total noise in all EM PPr channels in ADC counts, for the system equipped with (a) MCMs and (b) nMCMs.}
\label{fig:NoiseFullPpr}
\end{figure}
\begin{figure}[p!]
\begin{subfigure}[b]{0.49\textwidth}
\centering
\includegraphics[width=\textwidth]{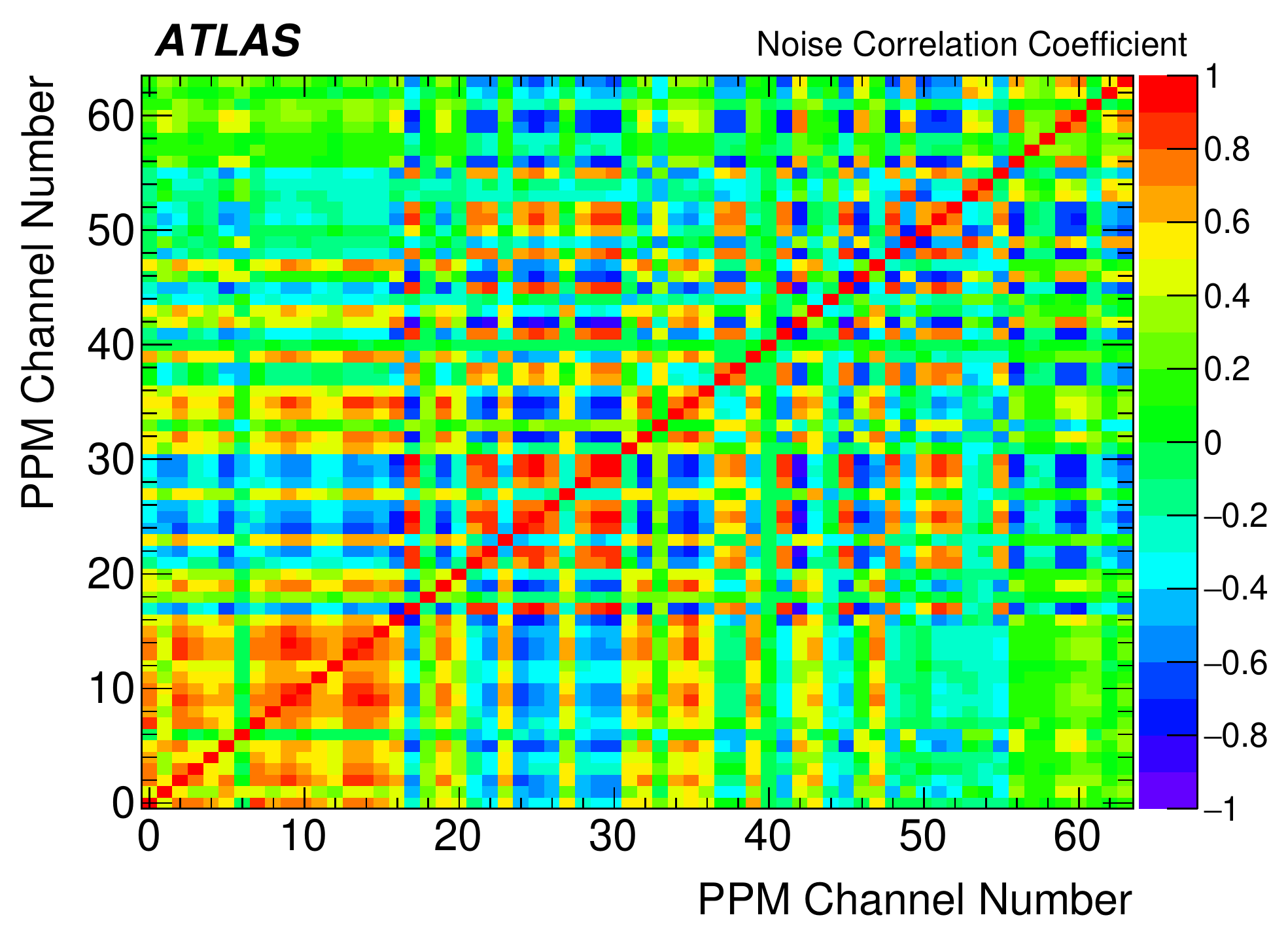}
\caption[]{}
\end{subfigure}
~
\begin{subfigure}[b]{0.49\textwidth}
\centering
\includegraphics[width=\textwidth]{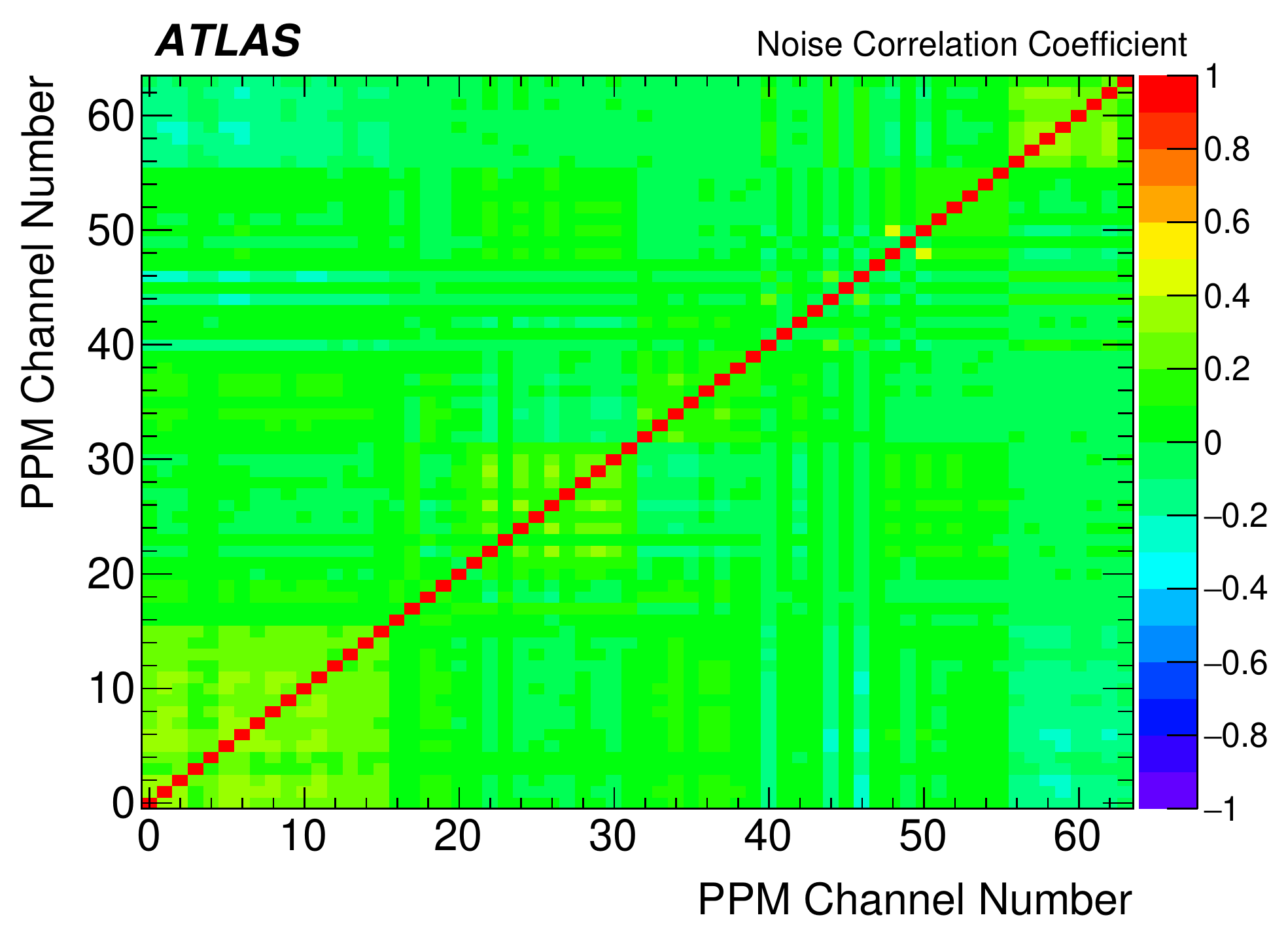}
\caption[]{}
\end{subfigure}
\caption[Correlated Noise]{Linear correlation coefficients as measured from noise for all possible pairs of channels on a PPM: (a) the result for a PPM equipped with MCMs, (b) the result for the same PPM equipped with nMCMs.}
\label{fig:NoiseCorrelation}
\end{figure}
 
On top of achieving a significant reduction, the noise became uniform across channels in the nMCM.
This is due to the different placement of the FADCs on the module.
On the MCM, two of the four FADCs are placed closer to the PPrASIC, such that the analogue signal path to them is longer than to the other two, making it more susceptible to crosstalk.
On the nMCM, both dual-channel FADCs are placed at the same distance from the CALIPPR FPGA, removing such differences.
 
A further feature of the noise on the MCM is its dependence on the fine-timing configuration, as shown in Figure~\ref{fig:NoisePpm}(b).
This is a consequence of the coupling between the analogue and digital ground planes on the PPM motherboard.
With the nMCM, this dependence is no longer observed, due to the reduced number of digital signals.
 
The impact of the improved nMCM noise on the PPr system as installed in the ATLAS detector is depicted in Figure~\ref{fig:NoiseFullPpr}.
It shows measurements of the noise for TTs in the EM calorimeter layer with (a) MCMs and (b) nMCMs.
Compared to the test-crate measurements in Figure~\ref{fig:NoisePpm}, the overall noise level in the real detector environment is
higher due to external contributions from the calorimeter detectors and front-end electronics, as well as from the analogue cables.
Compared to the central region, the noise is lower in the forward, high-$\eta$ regions, because the $E$ to $E_{\textrm{T}}$ conversion reduces the size of the external electronic noise.
For the MCM, the strong difference between channels already observed in Figure~\ref{fig:NoisePpm} is visible as a pattern of alternating high and low noise channels.
For the nMCM, this feature is no longer observed.
 
The correlation of the noise between channels on the same PPM is quantified by the linear correlation coefficient for all pairs of channels per PPM:
\begin{linenomath}
\begin{equation}
r_{AB} =  \frac {n\sum_{i=0}^n{A_i B_i} - \sum_{i=0}^n{A_i}\sum_{i=0}^n{B_i}}    {\sqrt{n\sum_{i=0}^n{A^2_i} - (\sum_{i=0}^n{A_i})^2}\sqrt{n\sum_{i=0}^n{B^2_i} - (\sum_{i=0}^n{B_i})^2}} \;, \notag
\label{equ:correlation}
\end{equation}
\end{linenomath}
where $A_i$ and $B_i$ are the ADC samples from two channels, and $n$ is the number of samples considered.
This measurement is performed on a per-module basis, with Receiver gain factors set to zero in order to minimise contributions from the calorimeter system.
The results for the MCM and the nMCM are shown in Figures~\ref{fig:NoiseCorrelation}(a) and \ref{fig:NoiseCorrelation}(b), respectively.
The larger scale structures that are visible come from groups of 16 neighbouring channels that share the same input cable and that are processed by the same AnIn board.
Smaller structures of four-by-four channels indicate the same MCM.
On the nMCM the correlation is still observed but is significantly smaller, indicating the removal of the common noise sources described above.

\subsection{Timing alignment}
\label{sec:performance_timing}
 
The L1Calo Trigger algorithms require all calorimeter information from the same BC to be processed synchronously.
This is achieved by two independent delay settings for coarse and fine timing adjustment of the analogue input signals.
 
Due to differences in cable length, the TT signals arrive at the PPr input with time differences of up
to 8~BCs, and thus have to be synchronised.
This is achieved by a `Sync.\ FIFO' (see Section~\ref{sec:nmcm_fw}), which delays the signals appropriately with BC precision.
 
Furthermore, the PPr algorithms require the digitisation of the analogue TT signal to occur as close in time as possible to the peak to ensure optimal energy resolution and BCID efficiency.
The exact quantitative condition depends on the pulse shape and thus varies by calorimeter region.
The extreme case is the EM FCAL, where a \SI{3}{\nano\second} deviation from the optimal sampling point would start to cause a degradation in energy response.
 
For precise adjustment, an additional fine timing is applied to each individual input signal, by delaying the ADC digitisation strobe in steps of about \SI{1.04}{\nano\second} relative to the LHC BC clock. Here the nMCM implementation provides slightly poorer resolution for the fine-timing configuration, with 24 steps of \SI{1.04}{\nano\second} within the BC as compared to the PHOS4 chip on the MCM, with 25 steps of \SI{1.00}{\nano\second}.
This difference does not lead to a loss in performance.
 
The individual delays are derived with a fit-based calibration procedure that uses digitised TT signals, which are recorded by the L1Calo readout system in $pp$ data taking.
Fitting optimised functions to the digitised pulses provides an estimate for the position of the maximum of the analogue input pulse.
The distance of this point from the time of digitisation gives the delay that has to be applied to the ADC strobe.
 
Due to the asymmetry of the pulse shapes and their dependency on the calorimeter regions (see Figure~\ref{fig:Pulses}), the convolution of either a Gaussian function and a Landau function (GLu) or two Landau functions (LLu) was found to describe the pulses best.
The GLu fit function is used in the calorimeter barrels, EMB and Tile, while pulses in the EM and HAD endcaps, as well as in the FCALs, are fitted with a LLu function.
A detailed description of the full method as applied in Run~1 is provided in Ref.~\cite{timing_note}.

This method, however, is limited by the sampling rate of the FADC.
In the Run~1 method, using the MCM with \SI{40}{\mega\hertz} sampling rate, i.e.\ \SI{25}{\nano\second}, five samples cover the positive part of the pulse relevant for the $E_{\textrm{T}}$ determination.
To handle the limited number of degrees of freedom in the fit, several fit parameters, like the pulse width, are fixed to values obtained by injecting known charges to calorimeter channels.
The slight difference in pulse shape between signals originating from artificial charge injection  and from real ionization by an incident particle is taken into account by $\eta$-dependent scaling factors for the pulse width.
 
In Run~2, the new FADCs on the nMCM sample the analogue input pulses at a rate of 80~MHz, i.e.\ twice the frequency of the FADCs on the MCM.
This way, the TT pulse shapes can be recorded with a higher resolution, typically up to nine samples.
This increased number of samples allows the fit to be performed almost freely, eliminating the necessity to use charge injection pulses in the calibration procedure.
Only one parameter, which determines the depth of the undershoot of the signal, remains fixed.
 
\begin{figure}[t]
\centering
\begin{subfigure}[b]{0.49\textwidth}
\centering
\includegraphics[width=\textwidth]{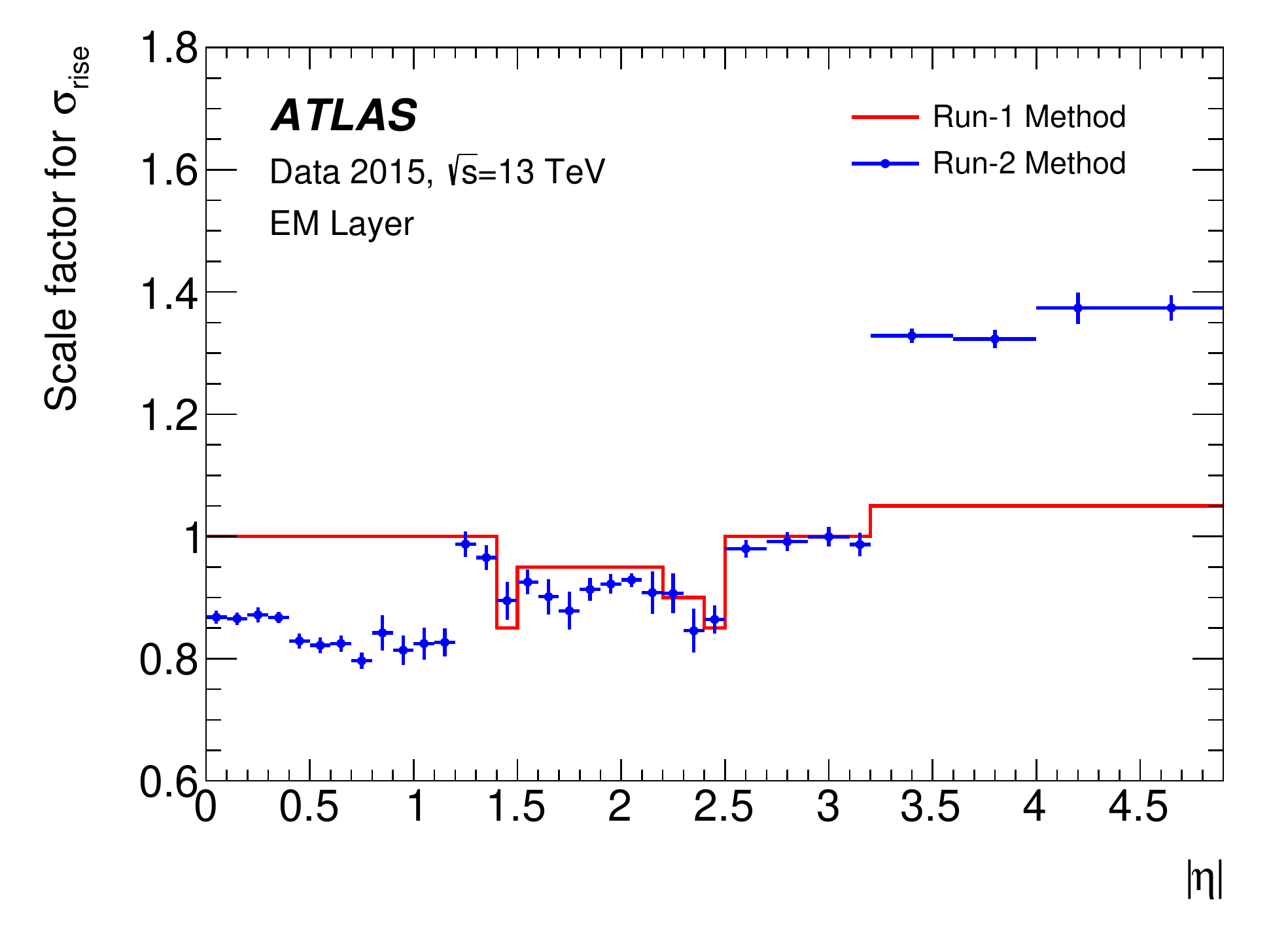}
\caption[]{}
\end{subfigure}
~
\begin{subfigure}[b]{0.49\textwidth}
\centering
\includegraphics[width=\textwidth]{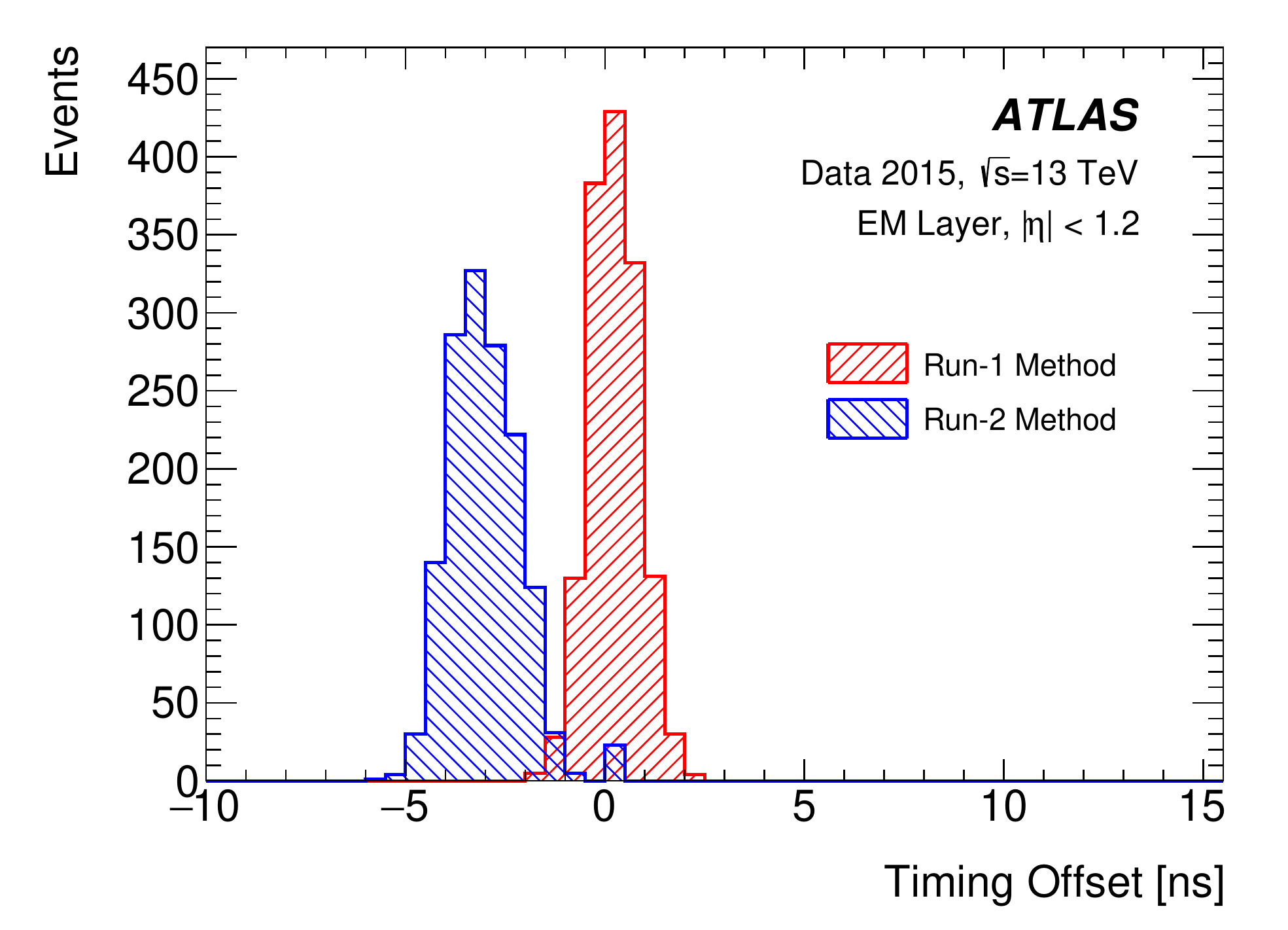}
\caption[]{}
\end{subfigure}
\caption[Input Timing]{(a) The scale factors for the width of the rising edge of the pulse, $\sigma_\textrm{rise}$, as function of $|\eta|$. (b) The different timing delay distributions obtained from fits to EM TT pulses in the very central $\eta$ region. Both plots compare the results from the Run~1 and Run~2 methods.}
\label{fig:TimingPerformance}
\end{figure}
 
The impact of this change on the calibration method is demonstrated in Figure~\ref{fig:TimingPerformance}.
Figure~\ref{fig:TimingPerformance}(a) shows the scale factors for the width of the rising edge of the pulse, $\sigma_\textrm{rise}$, as function of $\eta$.
Depending on the fit function used in the corresponding calorimeter region, $\sigma_\textrm{rise}$ denotes the width of either a Gaussian or a Landau function.
For the Run 1 method, where the pulse width in collision data is not determined directly due to limitations from the 40 MHz resolution, the figure shows the estimated scale factors used to correct the width of the charge injection signals.
For the Run 2 method, where the pulse widths are determined directly by the free fits to the collision data, the figure shows the ratio of the free fit results to the widths of the charge injection signals.
Especially in the EMB and FCAL regions, differences are revealed between the two methods, indicating potential improvements when updating the timing delays based on the Run~2 method.
Figure~\ref{fig:TimingPerformance}(b) shows the distribution of the resulting input timing delays for both the fits constrained by pulses recorded with the charge injection system, and the free fits to pulses recorded at 80~MHz, for TTs in the EM calorimeter in the region $|\eta| < 1.2$.
The change reveals a shift of \SI{4}{\nano\second} of the mean of the distribution, indicating an inaccuracy that was not observable with the Run~1 fine-timing procedure. However, no difference in trigger performance within the measurement accuracy was found after updating the timing delays based on the Run~2 method.
 
\subsection{Peak-Finder BCID}
\label{sec:performance_bcid_pf}
 
A central task of the PreProcessor system is the bunch crossing identification, BCID.
There are two distinctly different BCID methods:
\begin{itemize}
\item A Peak-Finder optimised for low-$E_{\textrm{T}}$ signals.
\item A threshold-based algorithm for pulses that saturate the ADCs.
\end{itemize}
 
This subsection covers the first method.
The Peak-Finder BCID utilises a noise filter, which is implemented as a FIR filter.
In the same way as for Run~1, the filter value $f_i$ is computed at each BC $i$ by convolving the last five recorded \SI{40}{\mega\hertz} ADC samples $A_{i-k}$ with five filter coefficients $c_k$:
\begin{linenomath}
\begin{equation}
f_i = \sum_{k=0}^{4} c_k\,A_{i-k}\;. \notag
\label{equ:filter}
\end{equation}
\end{linenomath}
On the nMCM, the new pedestal correction algorithm (see Section~\ref{sec:performance_pedcorr}) removes pile-up contributions from the filter output.
The resulting, corrected filter output for subsequent BCs is then checked for the peak condition:
\begin{linenomath}
\begin{equation}
f_{i-1}^{\,\textrm{corr}} < f_i^{\,\textrm{corr}} \geq f_{i+1}^{\,\textrm{corr}}\;. \notag
\label{equ:peakfinder}
\end{equation}
\end{linenomath}
For each BC that fulfils this condition, an $E_{\textrm{T}}$ value is obtained by processing the filter output through a LUT.
If the filter output is below a configurable noise cut, the LUT assigns zero $E_{\textrm{T}}$ (see Section~\ref{sec:performance_lut}).
The $E_{\textrm{T}}$ result is also set to zero for all BCs which fail the peak condition.
 
The performance of this BCID algorithm depends strongly on the chosen filter coefficients, which need to be optimised for the LHC operating conditions.
Both the LHC bunch filling scheme and the pile-up of the individual BCs have to be considered.
 
The nMCM provides more flexibility in assigning values to the filter coefficients.
For the MCM, the three central filter coefficients are 4-bit unsigned integers, while the outer two are 3-bit signed integers.
On the nMCM, all coefficients are 4-bit signed integer values.
This allows more-general filtering schemes to be used in order to mitigate the increased pile-up effects in Run~2.
 
The standard filter set-up in Run~1 was the `matched filter', which gives optimal performance when white noise is present.
These coefficients resemble the expected shape of the signal around the peak, with the consequence that genuine signal peaks experience a strong amplification while the noise is averaged out.
Because this filter works less reliably in high pile-up situations with strongly correlated background noise, impacting the performance of $E_{\textrm{T}}^{\textrm{miss}}$, multi-jet and low-energy L1 trigger items, a different approach was chosen for Run~2.
 
Autocorrelation (AC) filters use the self-correlation between nearby ADC samples in the absence of significant signals as additional information~\cite{clelandstern, ad_thesis}.
Dedicated runs in order to record such background samples were taken on a regular basis throughout Run~2 in order to optimise the BCID performance of the filters.
The five AC filter coefficients $c_i$ are calculated as
\begin{linenomath}
\begin{equation}
c_i = \sum_{j=0}^4 r_{ij}\,g_j\;, \notag
\label{equ:acfilter}
\end{equation}
\end{linenomath}
with the normalised pulse shape vector $\mathbf{g}$ and the autocorrelation matrix $\mathbf{r}$ given by
\begin{linenomath}
\begin{equation}
r_{ij} =  \frac{n\sum{A_i A_j} - \sum{A_i}\sum{A_j}} {\sqrt{n\sum{A^2_i} - (\sum{A_i})^2}\sqrt{n\sum{A^2_j} - (\sum{A_j})^2}} \;. \notag
\label{equ:acmatrix}
\end{equation}
\end{linenomath}
Here $A_i$ and $A_j$ denote two ADC samples of the same trigger tower within five consecutive BCs $i$ and $j$, respectively, and the sums run over the total number of recorded events $n$.
The normalised pulse shape vectors $\mathbf{g}$ for physics collisions had been measured during Run~1 using an oscilloscope in all different regions of the EM and HAD calorimeter (EMB, EMEC, FCAL1, Tile, HEC, FCAL23).
 
For low pile-up, where the correlation between ADC samples is low, $\mathbf{r}$ approaches the unity matrix and the AC filter coefficients become proportional to the pulse shape, reproducing the matched filters.
For high pile-up, the off-diagonal components of the autocorrelation matrix become more significant and the AC filters thus take a different shape, usually with large negative numbers for the off-centre coefficients.
 
\begin{figure}[t]
\centering
\begin{subfigure}[b]{0.49\textwidth}
\centering
\includegraphics[width=\textwidth]{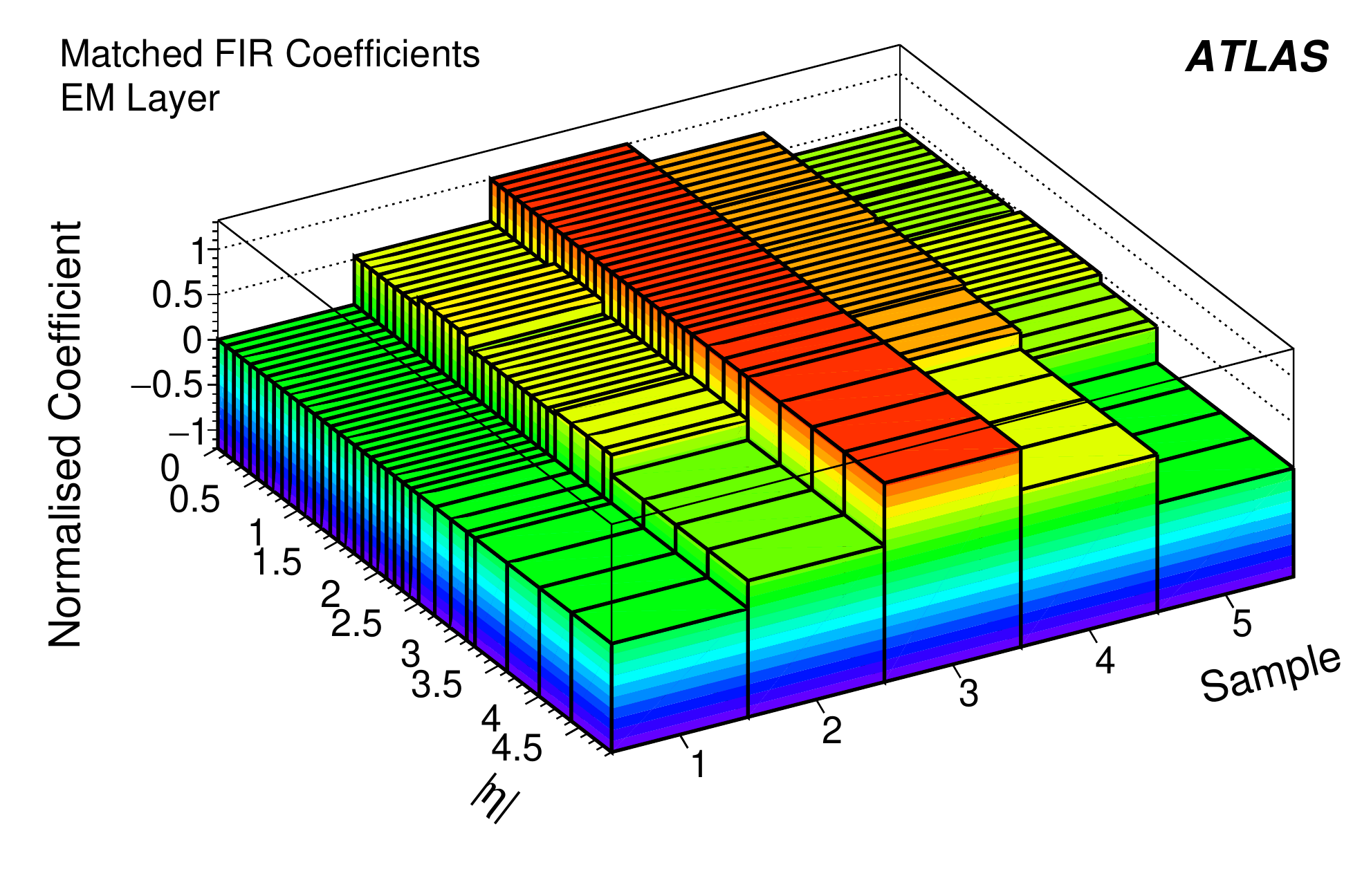}
\caption[]{}
\end{subfigure}
~
\begin{subfigure}[b]{0.49\textwidth}
\centering
\includegraphics[width=\textwidth]{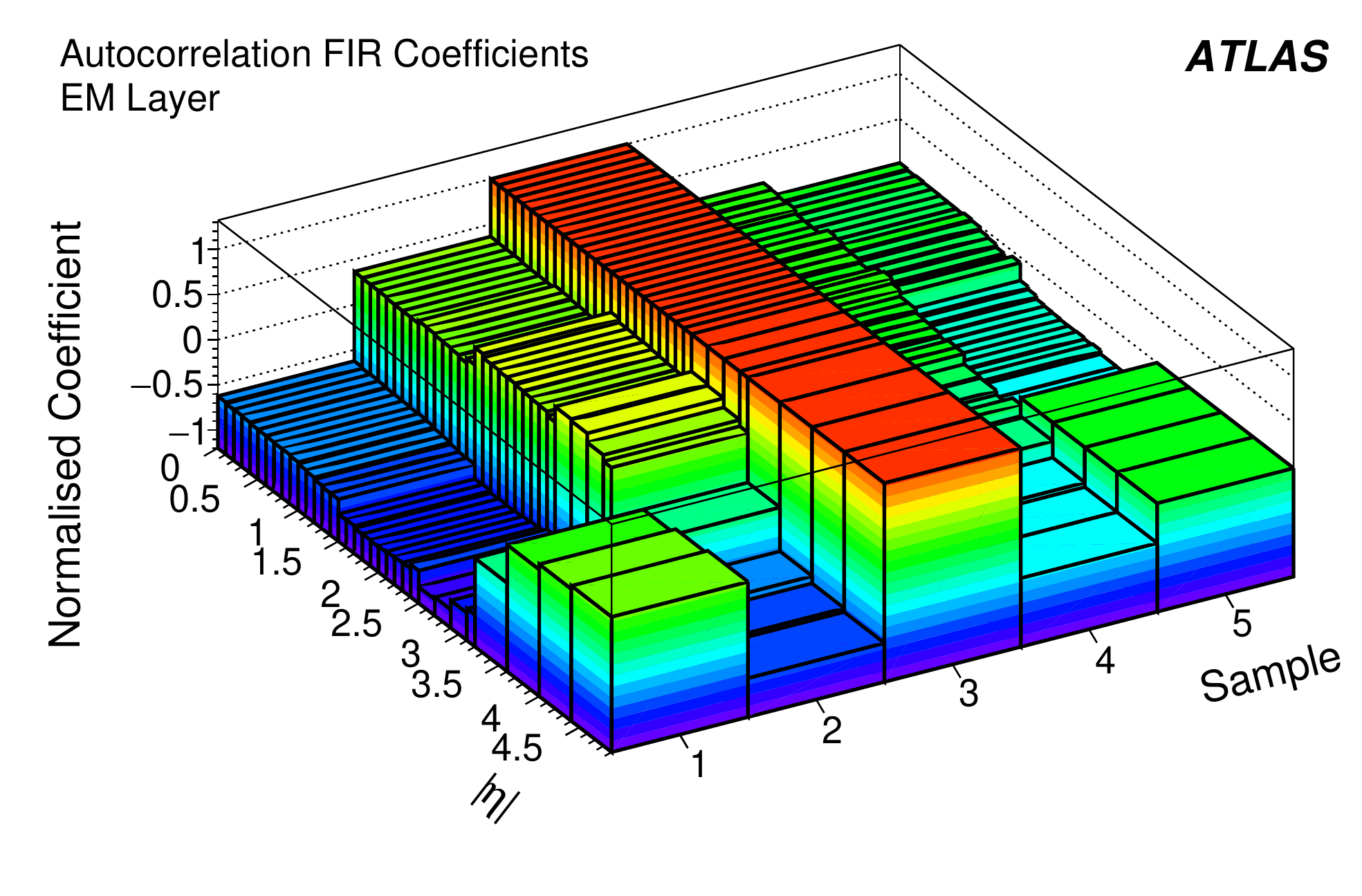}
\caption[]{}
\end{subfigure}
\caption[Matched and Autocorrelated Filter Coefficients]{Filter coefficient values for the EM calorimeter, normalised to the central coefficient, as a function of $|\eta|$: (a) the matched filter, (b) the AC filter scheme.}
\label{fig:FilterTypes}
\end{figure}

Figure~\ref{fig:FilterTypes} compares the coefficients of the matched filter with those of an AC filter derived from a data sample recorded in 2016 with an average number of $pp$ collisions per BC, $\langle\mu\rangle$, of 29.
The matched filter coefficients, shown in Figure~\ref{fig:FilterTypes}(a), are mostly homogeneous in $\eta$, with slight changes that reflect the different signal pulse shapes in the different parts of the calorimeter.
Figure~\ref{fig:FilterTypes}(b) shows the AC filters, which have a stronger dependence on $\eta$, especially towards the high-$\eta$ region where the off-centre and outer coefficients invert their behaviour.
This reflects the $\eta$ dependency of the pile-up, which is more pronounced in the forward region, due to the large size of the FCAL TTs.
The autocorrelation between a sample and its nearest neighbour is approximately \SI{60}{\percent} in the central EM barrel region, reaching up to \SI{80}{\percent} for the most forward FCAL TTs. With respect to the next-to-nearest neighbour, this correlation is approximately \SI{40}{\percent} in the central EM barrel region, and \SI{30}{\percent} for the most forward FCAL TTs.
The exact form of the AC matrices varies along $\eta$ , and depends on the LHC bunch filling scheme, as well as on the operational conditions.
 
\begin{figure}[p!]
\centering
\begin{subfigure}[b]{0.49\textwidth}
\centering
\includegraphics[width=\textwidth]{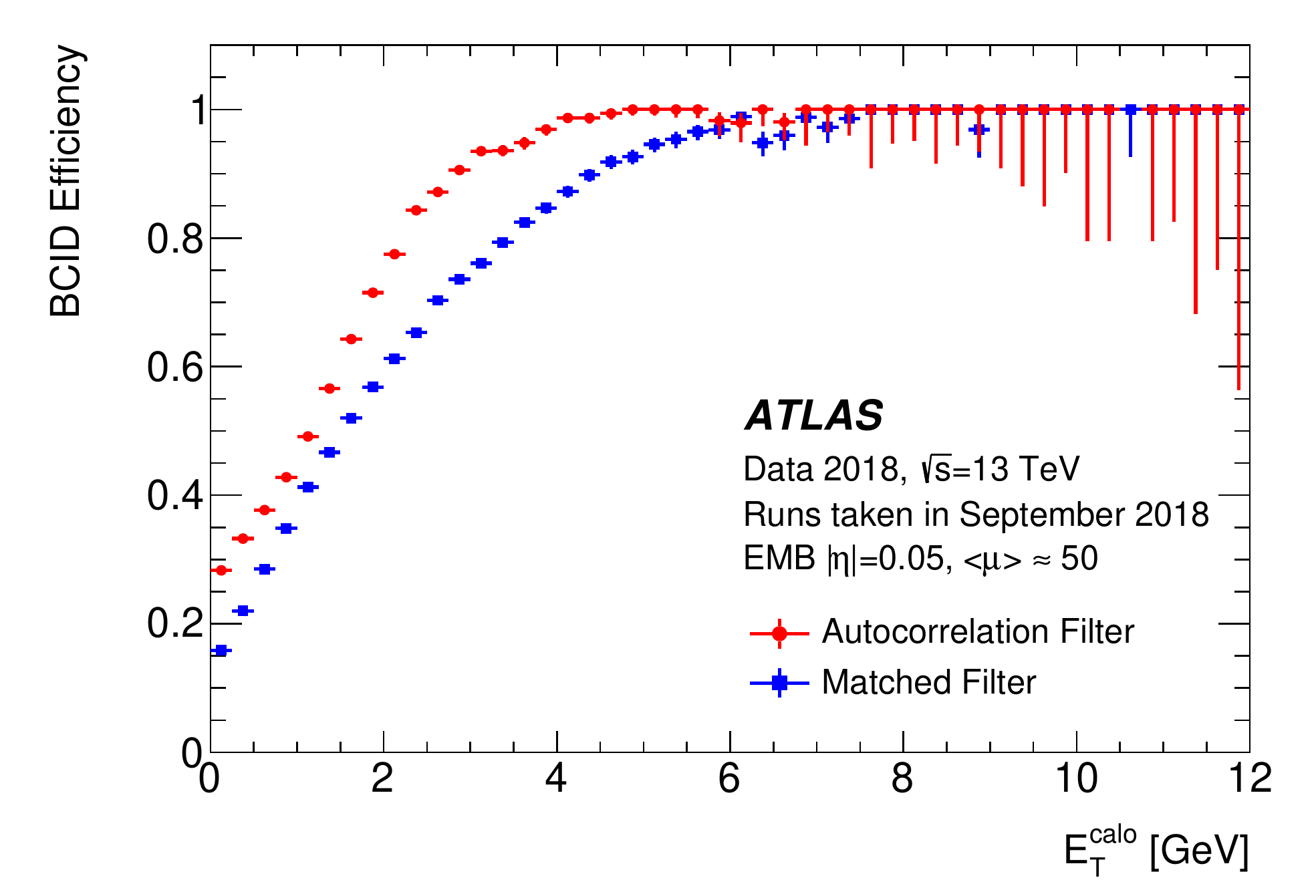}
\caption[]{}
\end{subfigure}
~
\begin{subfigure}[b]{0.49\textwidth}
\centering
\includegraphics[width=\textwidth]{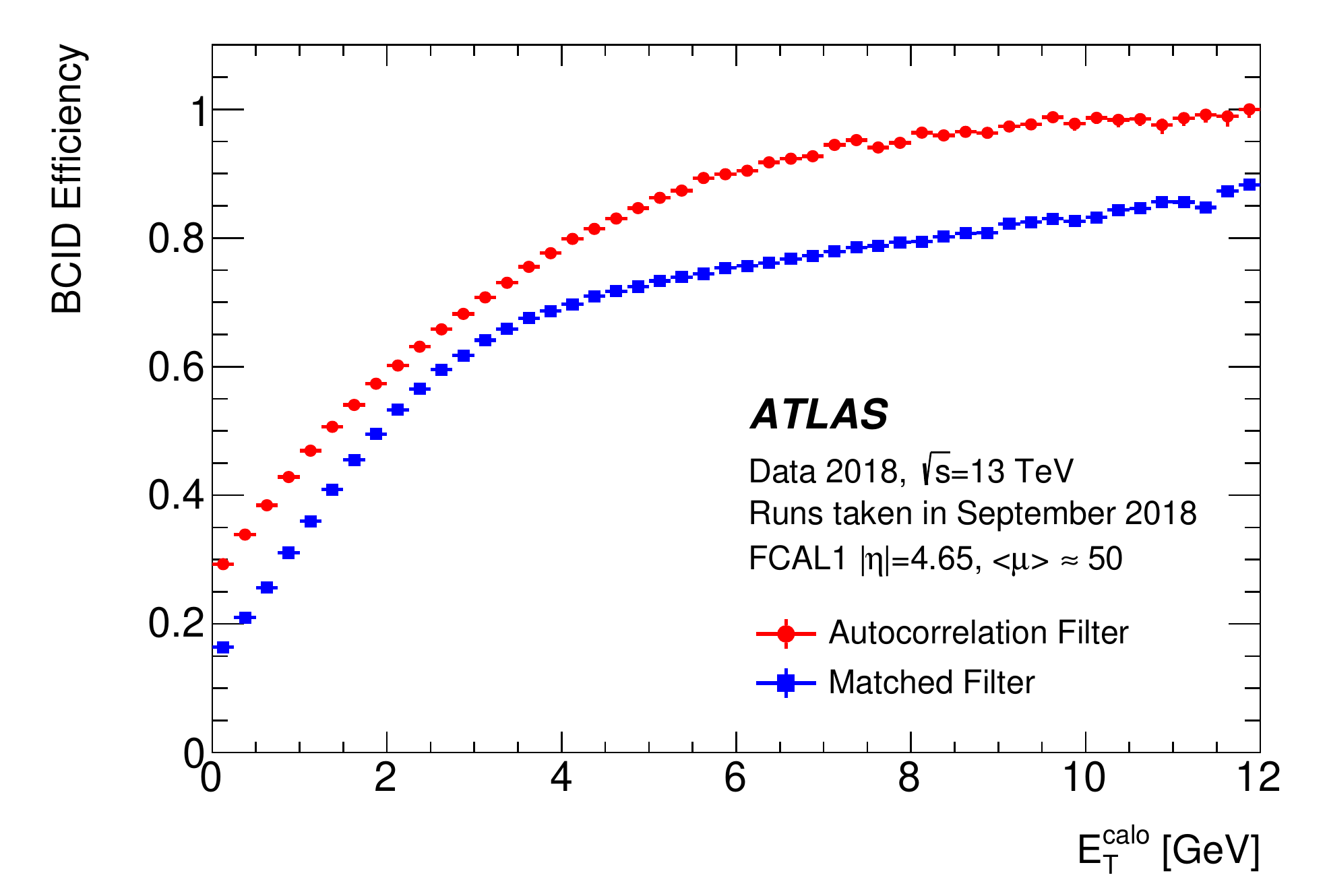}
\caption[]{}
\end{subfigure}
\caption[BCID Efficiency]{The efficiency of the Peak-Finder algorithm identifying the correct BC for both matched and AC filters in Run~2 as a function of transverse energy, for (a) the EM barrel and (b) the EM FCAL. The BCID efficiency is defined with respect to the transverse energy, $E_{\textrm{T}}^{\textrm{calo}}$, provided by the calorimeters for the respective BC.}
\label{fig:FilterEfficiency}
\end{figure}

Figure~\ref{fig:FilterEfficiency} shows the BCID efficiency as a function of the transverse energy measured by the calorimeters ($E_{\textrm{T}}^{\textrm{calo}}$) for both the matched and AC filters, as determined from 2018 data.
Randomly triggered events are used to provide a data sample with unbiased BCID decisions.
For each TT, $E_{\textrm{T}}^{\textrm{calo}}$ is calculated from the energy measurements of all calorimeter cells contained in the TT.
The BCID efficiency is then defined as the ratio of the number of events selected by the Peak-Finder algorithm to the number of events with given $E_{\textrm{T}}^{\textrm{calo}}$.
A significant background contribution is removed by discarding events which contain energy depositions that the calorimeter reconstruction does not assign to the analysed BC.
In the EM barrel region, shown in Figure~\ref{fig:FilterEfficiency}(a), the AC filter outperforms the matched filter at low energies and reaches the point of full efficiency earlier.
In the forward EM region, displayed in Figure~\ref{fig:FilterEfficiency}(b), this behaviour is even more pronounced.
 
\begin{figure}[p!]
\begin{subfigure}[b]{0.49\textwidth}
\centering
\includegraphics[width=\textwidth]{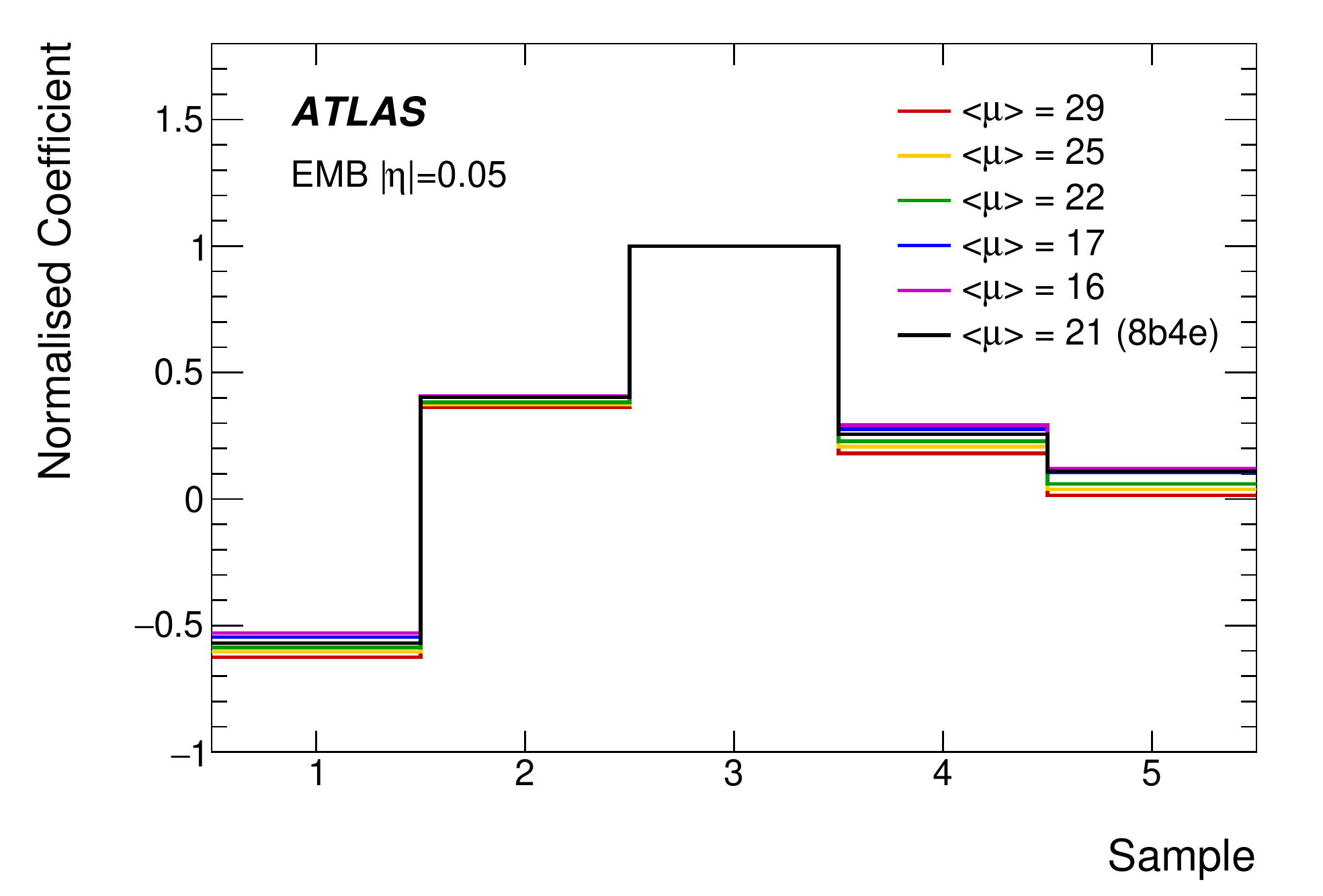}
\caption[]{}
\end{subfigure}
~
\begin{subfigure}[b]{0.49\textwidth}
\centering
\includegraphics[width=\textwidth]{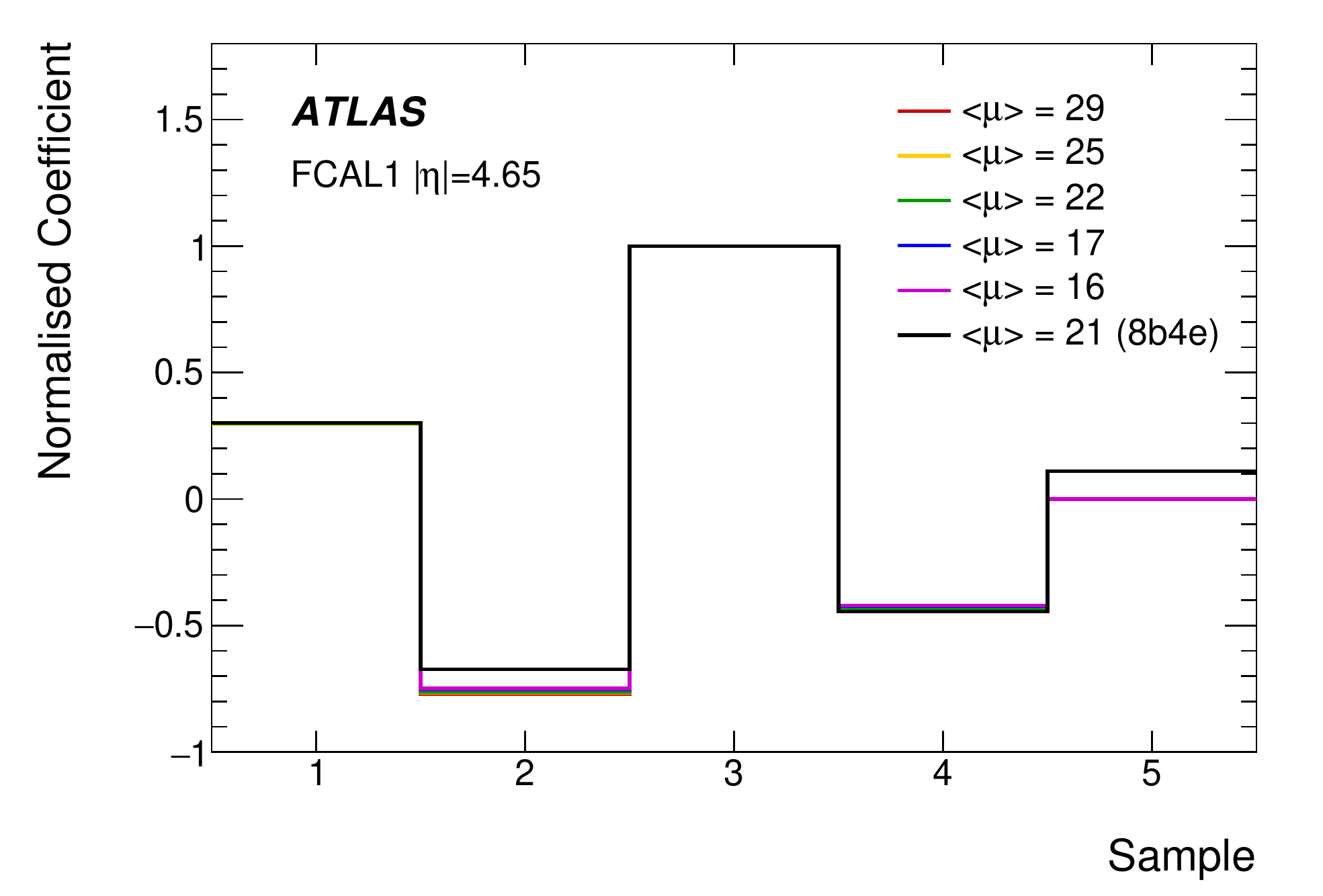}
\caption[]{}
\end{subfigure}
\caption[FIR Filter Pile-up Dependency]{Filter coefficient values, normalised to the central coefficient, for (a) TTs in the central $|\eta|$ region of the EMB and (b) TTs in the most forward $|\eta|$ region of FCAL1. Shown are the five coefficients derived from six different data samples with varying values of $\langle\mu\rangle$, one of which was recorded during LHC operation using the 8b4e filling scheme.}
\label{fig:FilterMuDep}
\end{figure}
 
While the coefficient values and the performance of the filters differ significantly between the matched and AC filter schemes, and show a strong dependency on $\eta$, the variation of the AC filter coefficients with $\langle\mu\rangle$ is fairly moderate.
This is demonstrated in Figure~\ref{fig:FilterMuDep}, which compares the normalised AC coefficients derived from five different data samples with varying values of $\langle\mu\rangle$ recorded during the same LHC fill in 2016, as well as from a data sample recorded during LHC operation using the so-called 8b4e filling scheme (see Section~\ref{sec:performance_pedcorr}).
The nMCM firmware supports only one set of filter coefficients per TT. Those are chosen to provide the best performance at the beginning of the LHC fills, where the  instantaneous luminosity is highest and the most extreme pile-up values are reached. Due to the weak dependency of the AC coefficients on $\langle\mu\rangle$, the trigger performance varies only marginally during a fill.
 
On the nMCM, the analogue input signals are digitised at \SI{80}{\mega\hertz}. While the main motivation for the increased sampling rate is the improvement of the Saturated BCID algorithm (see Section~\ref{sec:performance_bcid_sat}), simulation studies also showed modest improvements in the performance of the Peak-Finder BCID when operated at \SI{80}{\mega\hertz}~\cite{ad_thesis}. In order to ensure flawless functionality, it is a fundamental requirement that the L1Calo Trigger decision can be recalculated offline for every recorded ATLAS event. Operating the Peak-Finder BCID at \SI{80}{\mega\hertz} would require an increase of the data volume in the readout such that the maximum L1A rate could not be achieved any longer. It was therefore decided to retain the Peak-Finder operation at \SI{40}{\mega\hertz}.

\subsection{Dynamic pedestal correction}
\label{sec:performance_pedcorr}
 
The LHC machine configuration underwent a number of alterations during Run~2~\cite{LHC:Run2Rev}, optimising not only the instantaneous luminosity production but also affecting the detector configuration and performance.
In particular, two bunch filling schemes, defining which of the 3564 BC positions of the full LHC orbit are filled with protons, were used: the standard scheme based on long \SI{25}{\nano\second} bunch-trains, a consecutive series of BCs filled with protons, and the 8b4e filling scheme, used in the second half of 2017 to protect against effects from electron clouds in certain parts of the LHC ring.
In this scheme, the long trains are split into several shorter trains separated by very small gaps.
 
In the high pile-up environment of Run~2 the interplay between the bunch-train structure of the LHC beams and the long, bipolar calorimeter pulse shapes (cf.\ Figure~\ref{fig:Pulses}) results in localised distortions of the signal baseline (the pedestal).
Within a bunch-train, the overlap of pile-up hits from many subsequent BCs leads to an effective cancellation, as the positive peaks and the negative undershoots of overlapping signals on average yield a zero effect.
This is not the case at the start of bunch-trains, where only the positive contributions of pile-up hits are present, resulting in an effective rise of the pedestal value.
Similarly, the BCs at the end of a bunch-train show a net negative pile-up contribution, leading to a dip in the pedestal.
 
\begin{figure}[t]
\centering
\begin{subfigure}[b]{0.49\textwidth}
\centering
\includegraphics[width=\textwidth]{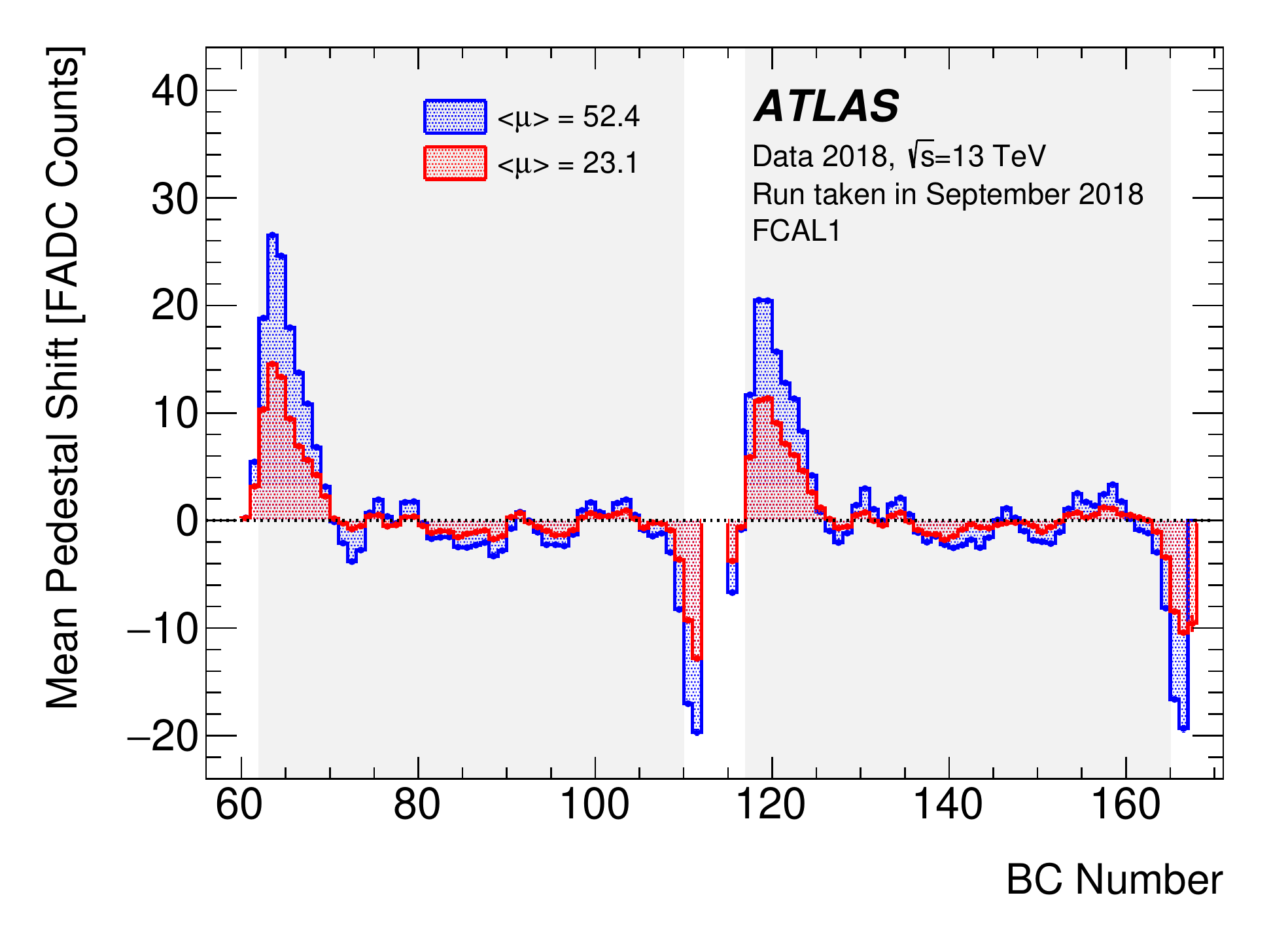}
\caption[]{}
\end{subfigure}
~
\begin{subfigure}[b]{0.49\textwidth}
\centering
\includegraphics[width=\textwidth]{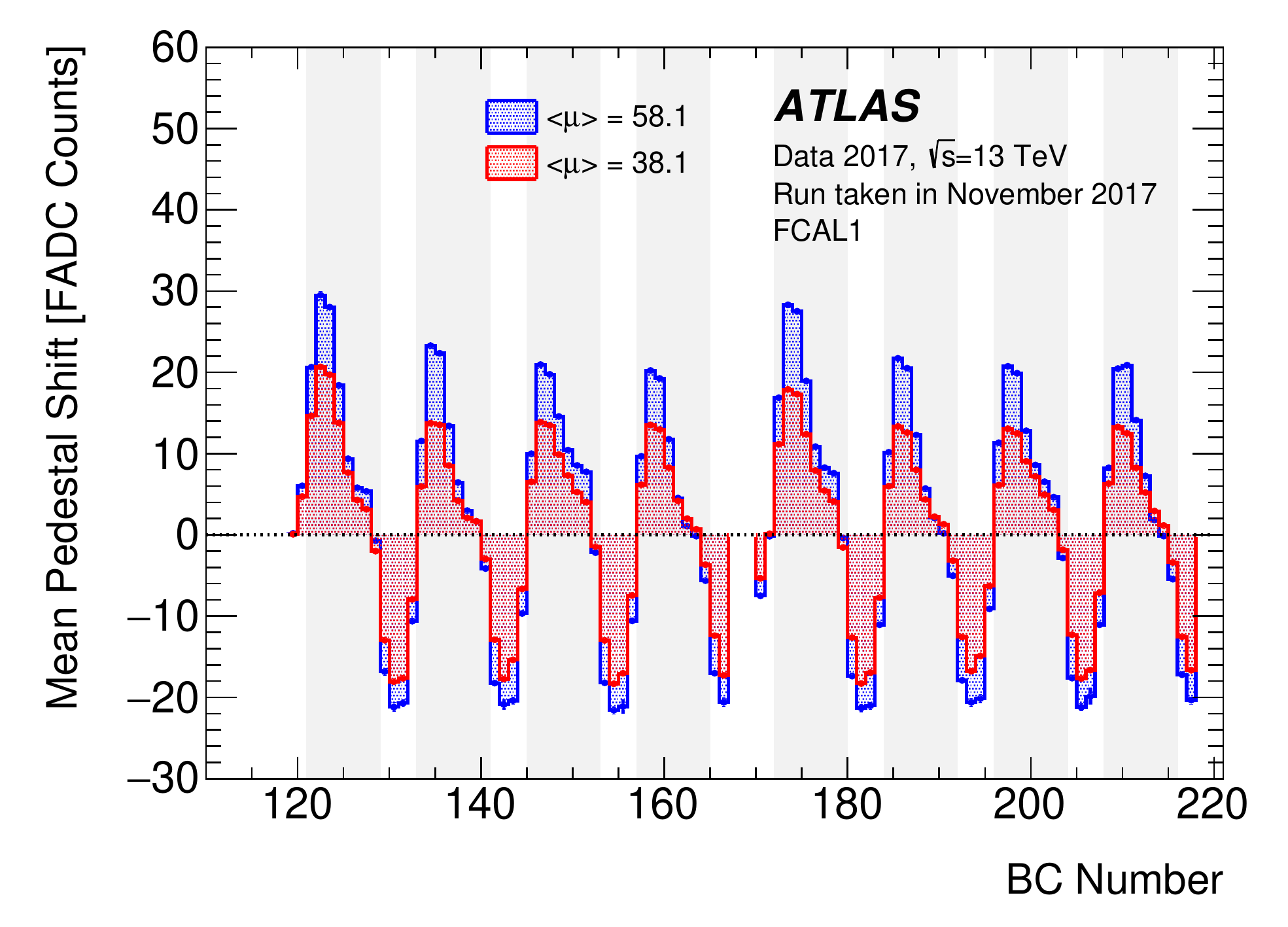}
\caption[]{}
\end{subfigure}
\caption[Pedestal Shift from Pile-Up]{The mean shift of the pedestal in the EM FCAL relative to its value in the absence of beams as a function of the BC Number in the LHC orbit. Results are shown for different $\langle\mu\rangle$ values, corresponding to different luminosity points in the same run. The shaded areas indicate BCs with colliding bunches in representative parts of the full LHC orbit of 3564 BCs. Results are shown for (a) an LHC filling scheme containing long bunch-trains and (b) an LHC fill using the 8b4e filling scheme.}
\label{fig:PedestalDrift}
\end{figure}

This behaviour of the pedestal is illustrated in Figure~\ref{fig:PedestalDrift} for the two different LHC filling schemes. Figure~\ref{fig:PedestalDrift}(a) shows the mean deviation of the pedestal from its default value for the standard filling scheme applied for most of Run~2, with two long bunch-trains separated by a short gap.
Figure~\ref{fig:PedestalDrift}(b) shows the pedestal shift for the  8b4e filling scheme.
In both cases, two distributions are presented, taken from different periods in the same ATLAS run, resulting in a different average number of $pp$ collisions per BC, $\langle\mu\rangle$, due to the decaying luminosity.
Since the 8b4e scheme results in many more BCs that show a significant pedestal shift, it produces a more challenging environment for the PPr algorithms.
 
\begin{figure}[t]
\centering
\includegraphics[width=0.60\textwidth]{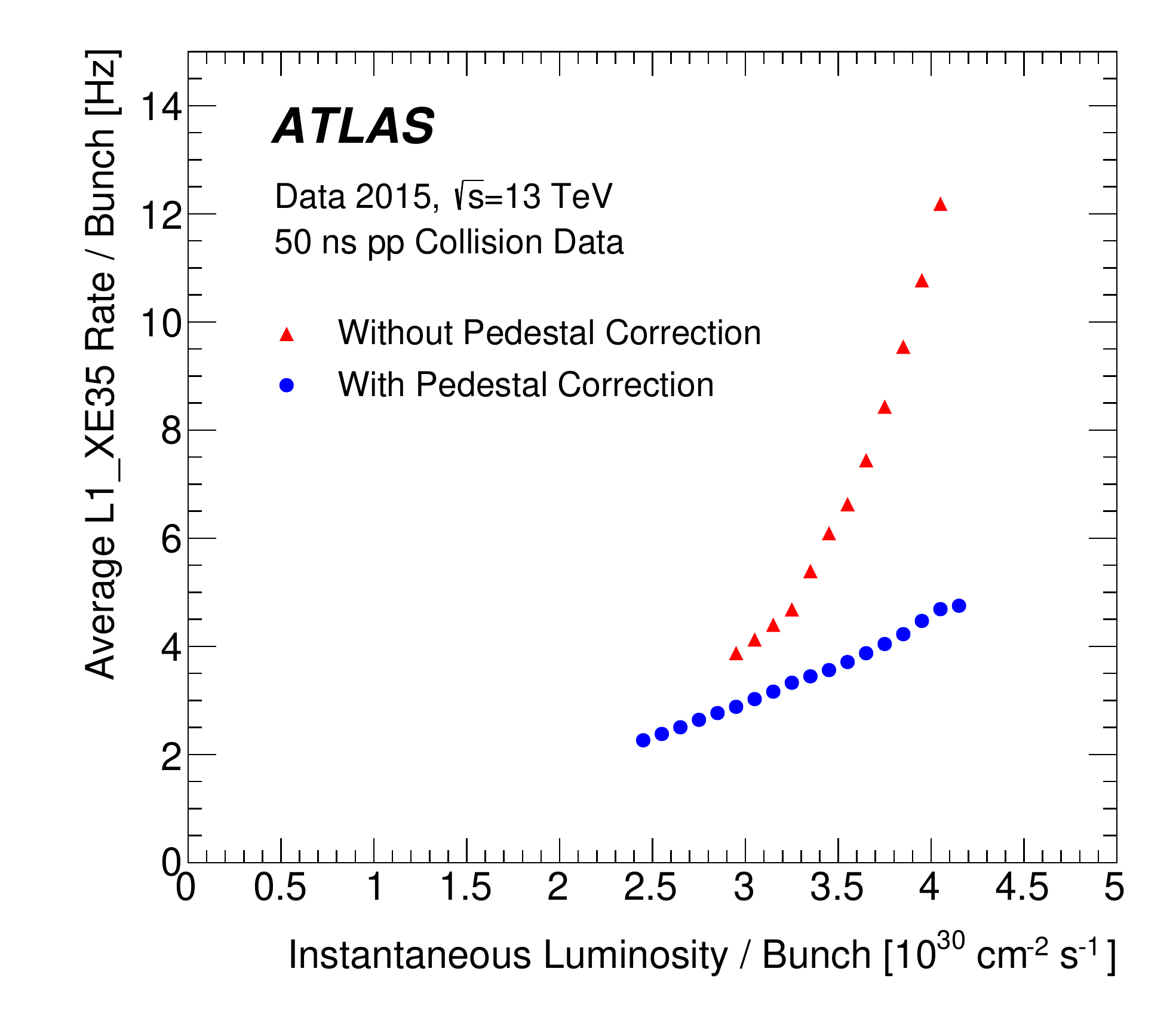}
\caption[Pedestal Correction Performance]{The rate of a L1 $E_{\textrm{T}}^{\textrm{miss}}$ trigger with a threshold of \SI{35}{\giga\electronvolt} as a function of instantaneous luminosity for two different runs at the beginning of 2015. One of the runs was taken without the pedestal correction, while the second run has the pedestal correction enabled, resulting in an almost linear behaviour of the rate. The efficiency of the $E_{\textrm{T}}^{\textrm{miss}}$ trigger is unaffected by this change. The figure is taken from Ref.~\cite{TRIG-2016-01}.}
\label{fig:pcvslumi}
\end{figure}

In the PPr signal processing, this pedestal shift results in a misinterpretation of the signal amplitude such that the deposited energy for signals in the affected BCs is overestimated.
In particular, noise spikes on top of this raised baseline can exceed the noise thresholds, leading to a non-zero $E_{\textrm{T}}$ result for affected TTs.
The impact of this effect varies between the different types of trigger objects identified by the L1Calo system, according to the number of TTs contributing to the reconstructed object.
While objects with narrow clusters, such as electrons, are barely affected by a small increase in energy, the missing transverse momentum, which is determined by summing the $E_{\textrm{T}}$ of all TTs in the calorimeter, shows non-linear L1 trigger rates as a function of luminosity (see Figure~\ref{fig:pcvslumi}).
For this reason, low-threshold L1 $E_{\textrm{T}}^{\textrm{miss}}$ triggers had to be disabled for the first 3 out of 36 filled bunch crossings of every bunch-train\footnote{The LHC operated with \SI{50}{\nano\second} bunch spacing schemes during Run~1.} at the beginning of the 2012 Run~1 data-taking period,
causing small losses of integrated luminosity for physics analysis using those triggers~\cite{ATL-DAQ-PUB-2018-001}.
 
A dedicated pedestal correction algorithm is implemented in the CALIPPR FPGA in order to reduce this effect, which otherwise could be mitigated only by increasing the noise cut thresholds.
The pedestal correction logic computes the average output of the FIR filter, independently for each of the 3564 BC positions of the full LHC orbit, and for each TT.
It then applies a correction such that the average of the resulting corrected filter output is equal to a configurable target value:
\begin{linenomath}
\begin{equation}
f_{i}^{\,\textrm{corr}} = f_{i} - \left( \bar{f_{i}} - f_{i}^\textrm{ped} \right) \;, \notag
\label{equ:pedcorr}
\end{equation}
\end{linenomath}
where $f_{i}^{\,\textrm{corr}}$ is the corrected filter output, $f_{i}$ is the original filter output, $\bar{f_{i}}$ is the average filter output calculated by the firmware, $f_{i}^\textrm{ped}$ is the configured target value and $i$ denotes the BC.
As the name implies, the target value $f_{i}^\textrm{ped}$ corresponds to the filter output for the expected flat pedestal in the absence of beam.
 
The average is recalculated at regular intervals, the length of which is given by the configurable number of orbits over which the average is taken.
This number has to be sufficiently large so that the averages are statistically stable, but at the same time small enough to allow the correction to follow the pile-up decrease with luminosity during an LHC run, and to minimise the impact of noise bursts in the calorimeters~\cite{LARG-2013-01}.
In Run~2 the default length is $2^{16}$ orbits, which corresponds to \SI{5.9}{\second} between updates of the averages.
This choice also ensures that the 32-bit-wide memories holding the sums of the 16-bit-wide filter output values will not overflow.
For longer intervals, the maximum length depends on the chosen filter coefficients and needs to be evaluated accordingly.
 
The positive effect of the pedestal correction on the trigger performance is illustrated in Figure~\ref{fig:pcvslumi}.
It shows the rate of an $E_{\textrm{T}}^{\textrm{miss}}$ trigger as function of the instantaneous luminosity for runs with and without the pedestal correction being enabled.
All quantities are normalised to the number of occupied bunches in the LHC, to accommodate different running conditions in the specific runs.
As seen, the pedestal shift induced by pile-up leads to a non-linear increase of the trigger rate with the luminosity.
This increase is cancelled out by the pedestal correction, resulting in the expected linear behaviour of the rate.
These runs were taken in the commissioning phase in 2015, at relatively low pile-up conditions compared to those achieved later in Run~2.
Producing a similar measurement for those very high pile-up conditions is not feasible, as operating the PPr without the pedestal correction would result in significant data loss due to high fake rates.
 
\subsection{BCID for saturated pulses}
\label{sec:performance_bcid_sat}
 
TT signals that exceed an $E_{\textrm{T}}$ of about \SI{250}{\giga\electronvolt} saturate the FADCs on the nMCM at 1023 ADC counts.
Any event containing a saturated signal is automatically accepted by the L1 trigger system, leaving the assignment to the correct BC as a remaining task.
 
At transverse energies above \SI{250}{\giga\electronvolt}, multiple ADC samples will stepwise saturate.
While the Peak-Finder algorithm remains fully efficient for a part of this energy range, it starts to fail once the deposited energy is above certain thresholds.
The exact $E_{\textrm{T}}$ value above which the Peak-Finder stops being fully efficient depends on the pulse shape of the particular TT.
A second BCID mechanism is thus implemented in the CALIPPR FPGA to identify the correct BC in case of multiple saturated ADC samples.
 
\subsubsection*{Saturated BCID algorithms}
 
The SatBCID algorithm compares the non-saturated ADC samples on the rising edge of the pulse with configurable thresholds.
The Run~1 SatBCID algorithm operated on the first two samples before saturation, and compared them with two thresholds:
\begin{itemize}
\item If $A[s-1] > H$ and $A[s-2] > L$, then identify $s$ as the correct BC,
\item otherwise, identify $s+1$ as the correct BC.
\end{itemize}
Here, $H$ and $L$ denote a high and low threshold, respectively, $A[k]$ denotes the value of the ADC sample at BC $k$, $s$ is the BC of the first saturated sample and the integer increments represent neighbouring BCs relative to $s$.
 
Due to complications in the threshold calibration procedure, the SatBCID algorithm was operated in a trivial configuration for most of Run~1.
In this configuration, the SatBCID algorithm always identifies the sample $s+1$, and is operated in parallel with the Peak-Finder algorithm that correctly identifies all pulses where $s$ is the peak sample.
The use of either of these algorithms can be determined by the BCID Decision Logic as described in the second of the following subsections.
This set-up was validated as identifying the correct sample over the whole LHC energy range accessible in Run~1.\footnote{
In Run~1, energies up to $\sqrt{s} = \SI{8}{\tera\electronvolt}$ were reached, while in Run~2 the LHC achieved energies up to \SI{13}{\tera\electronvolt}.
Due to momentum conservation, at most half of this can be deposited in a localised region.}
 
\begin{figure}[t]
\centering
\includegraphics[width=0.9\textwidth]{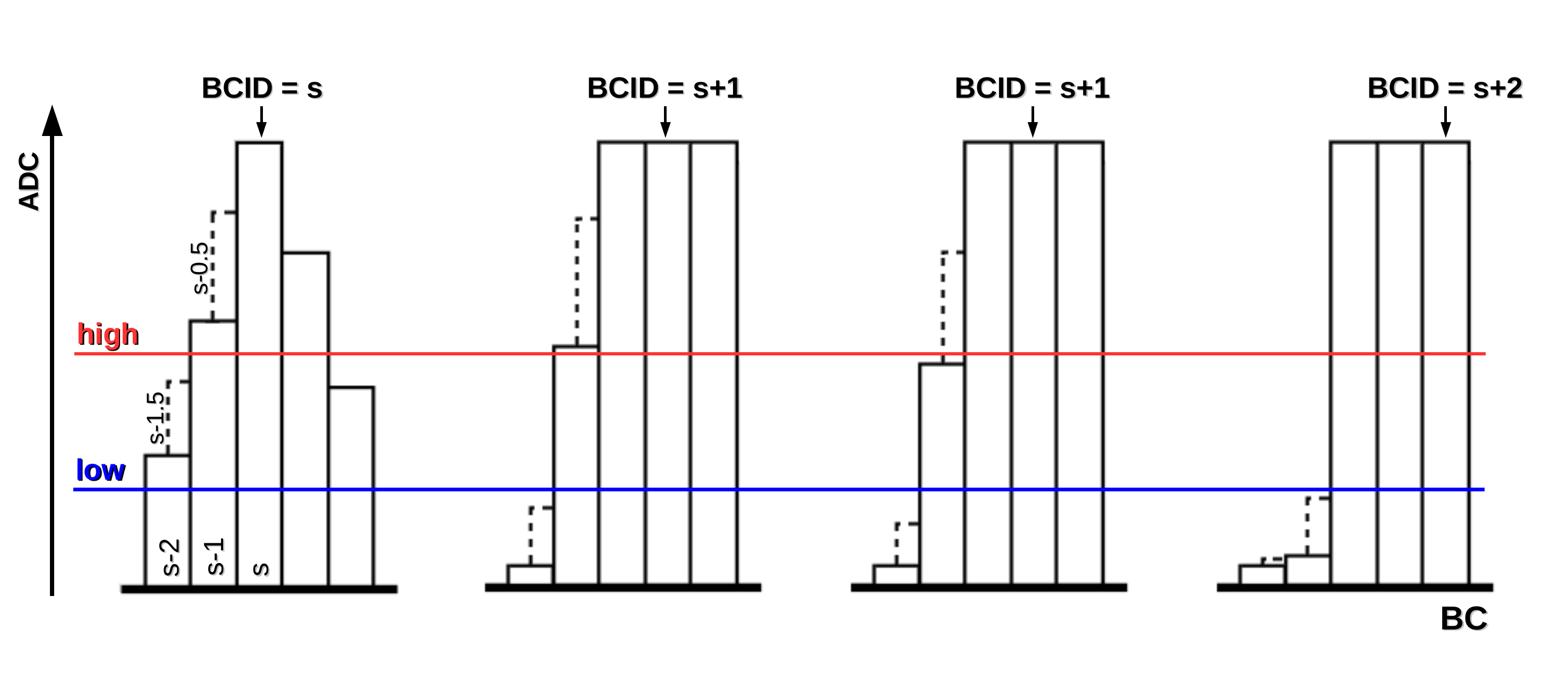}
\caption[Sat80BCID Algorithm]{Illustration of the four cases of the Sat80BCID algorithm. Solid lines represent samples digitised at \SI{40}{\mega\hertz}, while the dashed lines are additional samples digitised at \SI{80}{\mega\hertz}. For the \SI{80}{\mega\hertz} samples, only those used in the algorithm are shown.}
\label{fig:SatBcidAlgo}
\end{figure}

On the nMCM, the SatBCID algorithm was updated to gain additional flexibility for the calibration procedure and to extend the validated range to the full Run~2 energy.
The updated version uses the additional ADC samples acquired by the \SI{80}{\mega\hertz} FADCs installed on the nMCM.
Using the same notation as above, this Sat80BCID algorithm is:
\begin{itemize}
\item If $A[s-1] > H$ and $A[s-1.5] > L$, then identify $s$ as the correct BC,
\item if $A[s-1] > H$ and $A[s-1.5] \leq L$, then identify $s+1$ as the correct BC,
\item if $A[s-1] \leq H$ and $A[s-0.5] > H$, then identify $s+1$ as the correct BC,
\item if $A[s-1] \leq H$ and $A[s-0.5] \leq H$, then identify $s+2$ as the correct BC.
\end{itemize}
The fractional sample identifiers refer to the additional \SI{80}{\mega\hertz} samples.
The four cases of the algorithm are shown in Figure~\ref{fig:SatBcidAlgo}.
 
The sample $s-2$ used in the original SatBCID algorithm is situated far from the peak of the pulse and typically behaves non-linearly as a function of $E_{\textrm{T}}$.
By instead choosing the $s-1.5$ sample, which is closer to the peak, this effect is reduced and the reliability of the Sat80BCID algorithm is increased.
 
The very high energy regime, where multiple samples in the pulse saturate, is covered by introducing the third sample, $s-0.5$, in the algorithm.
This allows pulses with two saturated samples before the peak to also be identified correctly, which was not possible with the Run~1 algorithm.
 
\subsubsection*{Calibration of the Sat80BCID algorithm}
The calibration of the Sat80BCID algorithm consists of finding values for the two thresholds $H$ and $L$ such that the logic outlined above always selects the correct peak sample.
Imposing this requirement on the algorithm logic provides four bounds for the thresholds:
\begin{itemize}
\item $A[n-2.5]$@$E_{\textrm{T}}^{\textrm{sat}}[n-2] < L < A[n-1.5]$@$E_{\textrm{T}}^{\textrm{sat}}[n]$,
\item $H > \min\left( A[n-1] @ E_{\textrm{T}}^{\textrm{sat}}[n], \enskip A[n-1.5] @ E_{\textrm{T}}^{\textrm{sat}}[n-1] \right)$.
\end{itemize}
Here, $n$ denotes the peak sample of the pulse, and $A[i]$@$E_{\textrm{T}}^{\textrm{sat}}[k]$ is the height of the ADC sample for BC $i$ at that transverse energy $E_{\textrm{T}}^{\textrm{sat}}$ for which the sample at BC $k$ is the first one to saturate the ADC.
 
\begin{figure}[t]
\centering
\begin{subfigure}[b]{0.49\textwidth}
\centering
\includegraphics[width=\textwidth]{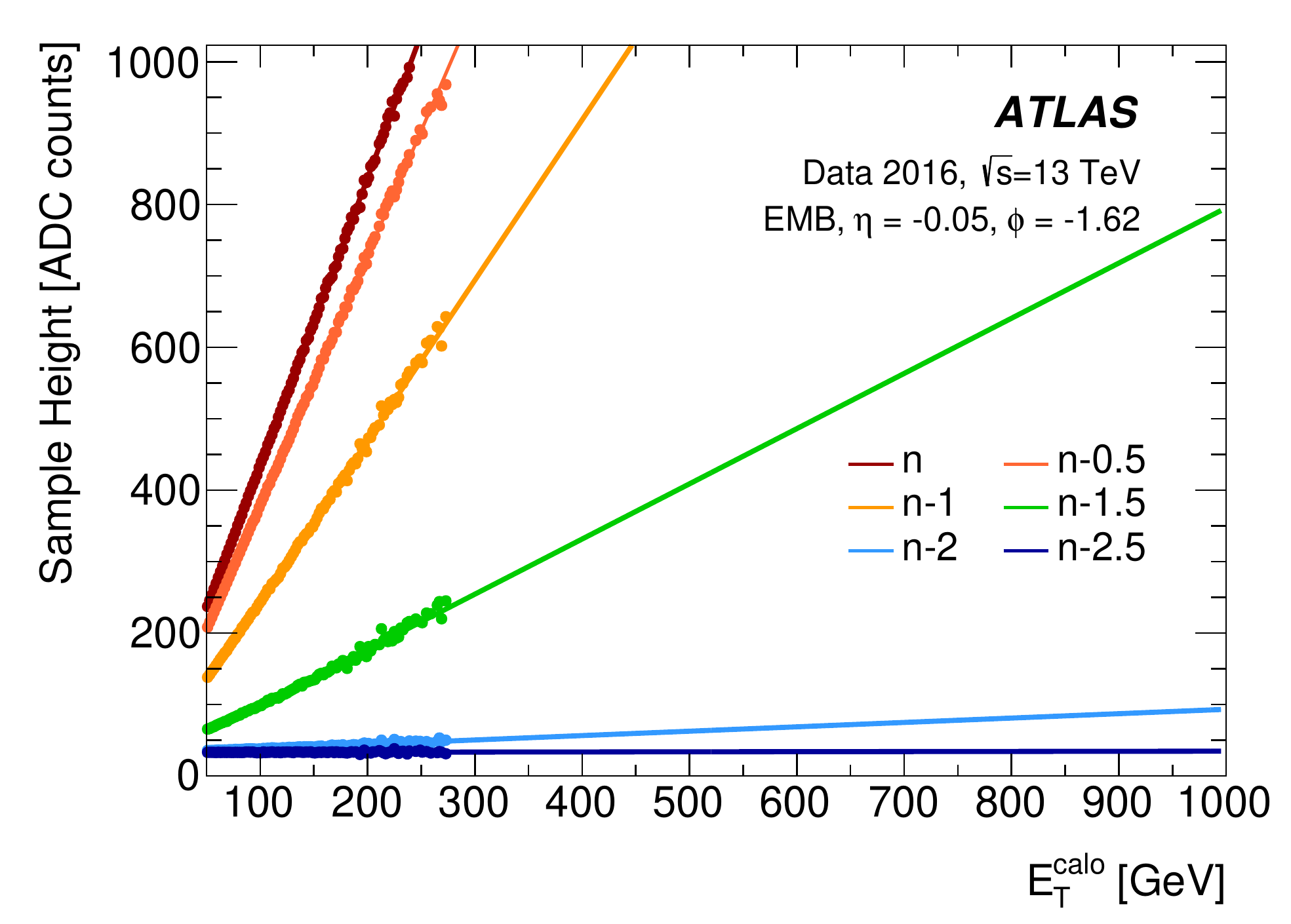}
\caption[]{}
\end{subfigure}
~
\begin{subfigure}[b]{0.49\textwidth}
\centering
\includegraphics[width=\textwidth]{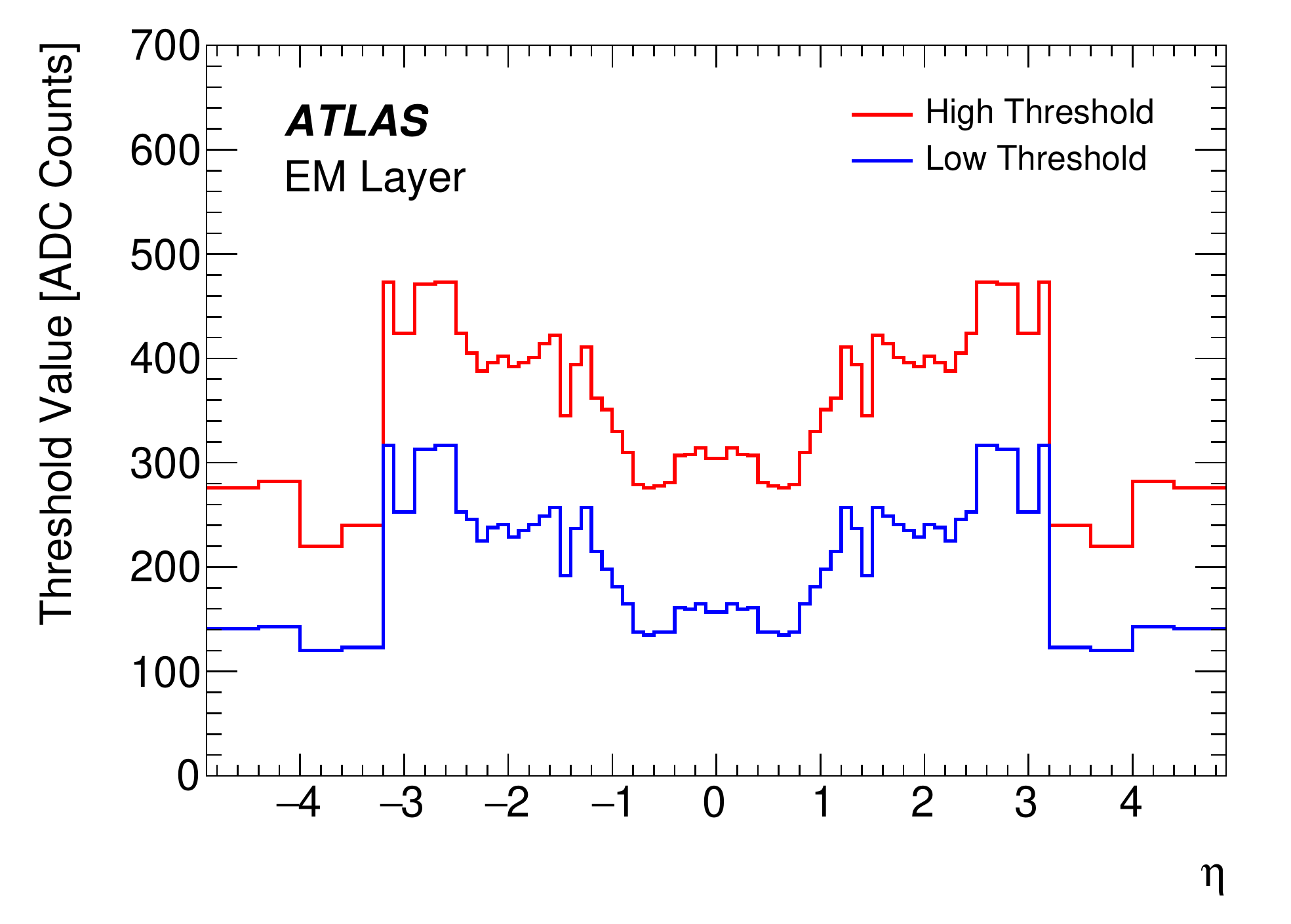}
\caption[]{}
\end{subfigure}
\caption[Sat80BCID Calibration]{Calibration of the Sat80BCID algorithm: (a) the behaviour of the ADC samples from the rising edge of pulses in an EMB TT as function of $E_{\textrm{T}}$ as measured in the calorimeters. The results of linear fits for each of the samples are indicated by lines of the respective colour. (b) The threshold values used in the EM calorimeter.}
\label{fig:sat80calibration}
\end{figure}

In order to obtain sufficient statistics to find these bounds from $pp$ collision data, the analysis is performed mainly on non-saturated pulses that are extrapolated to the saturated regime.
For this purpose, the ADC samples on the rising edge of the pulse are studied as a function of $E_{\textrm{T}}$, typically showing a linear behaviour.
This is illustrated in Figure~\ref{fig:sat80calibration}(a), which shows the $E_{\textrm{T}}$-behaviour of ADC samples for a TT in the EMB calorimeter.
 
Once the intervals are identified, the threshold values are set 15 to 20 ADC counts away from them, to account for noise and pile-up contributions.
Figure~\ref{fig:sat80calibration}(b) shows the resulting threshold values for all $\eta$-bins in the EM calorimeter.
These thresholds were used to operate the Sat80BCID algorithm for most of Run~2.

\subsubsection*{BCID Decision Logic}
\label{sec:performance_bcid_declog}
 
While the Peak-Finder and SatBCID algorithms are designed and optimised for separate energy regimes, both operate in parallel on all pulses.
In cases where the algorithms do not identify the same BC, the BCID Decision Logic determines which of the algorithm decisions is used.
 
In Run~1, the BCID Decision Logic was configured to select, for all energy regimes, the earlier BCID decision.
This functionality was used to facilitate the trivial set-up for saturated pulses in Run~1 that was described above, with the SatBCID algorithm by construction always selecting the $s+1$ sample.
For the case of matched filters, the Peak-Finder result moves to later BCs as more ADC samples saturate.
Therefore the BCID Decision Logic selects always the correct BCID provided that the Peak-Finder correctly identifies all pulses for which $s$ is the peak sample.
This set-up fundamentally relies on there being no saturated pulses for which $s+2$ is the correct choice.
The validity of this assumption was demonstrated for all energies accessible in Run~1.
 
This set-up breaks down for the AC filter in Run~2, as the Peak-Finder identifies earlier BCs as more ADC samples saturate.
As a result, the BCID Decision Logic selects the wrong Peak-Finder result instead of the correct SatBCID result.
This can lead to events being triggered one BC early, resulting in an almost complete loss of detector information for that event.
This is due to a dead-time of 4 BCs that is applied by the CTP after each accepted event in order to protect the data integrity during the readout process.
An approximate offline reconstruction can be attempted if the early triggered event was identified as such, for those detector components which read out information for more than the triggered BC. This, however, is not the case for the tracking detectors.
The opposite case, in which the $E_{\textrm{T}}$ result of a single trigger tower is falsely assigned to a later BC is much
less dangerous, as there is usually enough activity in the event to still cause a trigger decision in the correct BC.
 
For this reason, the CALIPPR implementation of the BCID Decision Logic introduces a new pulse height criterion, by also counting the number of saturated samples in the pulse.
The number of saturated ADC samples is compared with two thresholds, one for the Peak-Finder algorithm and the other for the SatBCID algorithm.
The Peak-Finder result is ignored if the corresponding threshold is exceeded, while the SatBCID result is ignored if the corresponding threshold is not reached.
The inefficiency of the AC filters for pulses with higher degrees of saturation is then avoided by ignoring the Peak-Finder decision at those energies.
In the configuration used during most of Run~2, the Peak-Finder is ignored for pulses with more than two saturated samples in the LAr calorimeter, and for pulses with more than three saturated samples in the Tile calorimeter.

\subsubsection*{Saturated BCID performance}
\label{subsec:satbcidperform}
 
The BCID treatment for saturated pulses underwent several changes at the start of Run~2 before reaching the set-up described above, leading to gradual improvements of the performance.
A summary of this is presented in Table~\ref{tab:sat40vs80}.
It shows the inefficiency of the BCID for saturated pulses in the different years and BCID set-ups during Run~2, expressed as the number of early triggered events.
These are events where the BCID algorithms assigned the energy of a saturated TT to the BC before the actual event.
Details of how to extract those events from the data are given in the following subsection addressing Saturated BCID monitoring.
Because the L1 trigger automatically selects events containing a saturated TT, such events result in an early trigger and thus the loss of the detector readout for the correct BC.
 
\begin{table}[t]
\centering
\caption[Saturated BCID Algorithm Performance]{Performance of the BCID for saturated pulses during different periods in Run~2. Each period is given by a year of data taking during Run~2, the algorithm set-up that was used and the integrated online luminosity $\mathcal{L}$ recorded by ATLAS during this time~\cite{ATLAS:2019pzw}. The performance is quantified by the rate of early triggered events which are caused by the L1Calo BCID algorithms misidentifying a saturated pulse.}
\vspace{3mm}
\begin{tabular}{|l|l|c|c|}
\hline
Year & SatBCID set-up & $\mathcal{L}$ [\si{\per\femto\barn}] & Early triggered events [per \si{\per\femto\barn}] \\
\hline
2015 & Run~1 set-up         & ~~3.9  & 30~~~~~   \\
2016 & New Decision Logic  & 31.1 & 1.3  \\
& Sat80BCID algorithm & ~~4.5  & 0~~~    \\
2017 & Sat80BCID algorithm & 46.9 & 0~~~    \\
2018 & Sat80BCID algorithm & 60.6 & 0.3  \\
\hline
\end{tabular}
\label{tab:sat40vs80}
\end{table}
 
At the start of Run~2, the Sat80BCID algorithm was operated in monitoring mode, such that its BCID decision was included in the event readout but not considered by the Decision Logic.
The Decision Logic was still configured in the Run~1 set-up.
This set-up led to the aforementioned problems for the AC filters in the saturated regime, causing a problematic rate of early triggered events.
 
In 2016, the new BCID Decision Logic was deployed, allowing the Peak-Finder result to be ignored for TTs with multiple saturated slices, which strongly reduced the rate of early triggered events.
Towards the end of that year, the Sat80BCID algorithm was activated, reducing the early BCID rate to zero for the rest of 2016 and also for all of 2017.
 
In 2018, 20 early triggered events were observed.
For these events, the combination of the chosen AC filter coefficients pushed the Peak-Finder result early for signals in a very small window of moderate energy ranges.
This lead to a small inefficiency in the Peak-Finder algorithm at energies just below where the Decision Logic ignores its result.

\subsubsection*{Saturated BCID monitoring}
\label{subsec:satbcidmoni}
 
Since the beginning of 2016, incorrectly timed candidate events were investigated systematically. This monitoring was based on dedicated triggers filtering the events for every run, and stepwise improved in the course of Run~2.
In 2018, the BCID performance for saturated pulses was monitored with a tool that automatically investigates each run immediately after its completion.
It analyses data from a specialised recording stream that contains events for which a BC directly before or after the L1A would have triggered
a highly energetic jet trigger item, L1\_J400, which requires a jet at L1 with transverse energy above approximately \SI{400}{\giga\electronvolt}.
While this is a typical signature of events with the wrong BCID decision involving saturated pulses, most contributions to this data stream originate from other sources, such as calorimeter noise bursts, pile-up coincidences, or cosmic-ray showers.
The monitoring tool removes such events by demanding a high-energy jet trigger in the selected BC (L1\_J100) and by applying various selection criteria to the pulse shapes and the number of TTs with significant $E_{\textrm{T}}$ depositions in all involved BCs.
For the remaining events, a set of plots that show the $E_{\textrm{T}}$ distribution in the calorimeters are prepared and any suspect TTs are indicated.
These plots are then checked by L1Calo Trigger operators, who perform the final categorisation based on the detailed properties of each individual event.
 
The monitoring evolved alongside the algorithmic implementation, starting from a first rudimentary, manually executed implementation in 2015. 
In 2016, the aforementioned specialised data stream was introduced, and the analysis was further refined and adapted to this data format also during 2017 and up to the automation in 2018.
 
\begin{figure}[p!]
\centering
\begin{subfigure}[b]{0.82\textwidth}
\centering
\includegraphics[width=\textwidth]{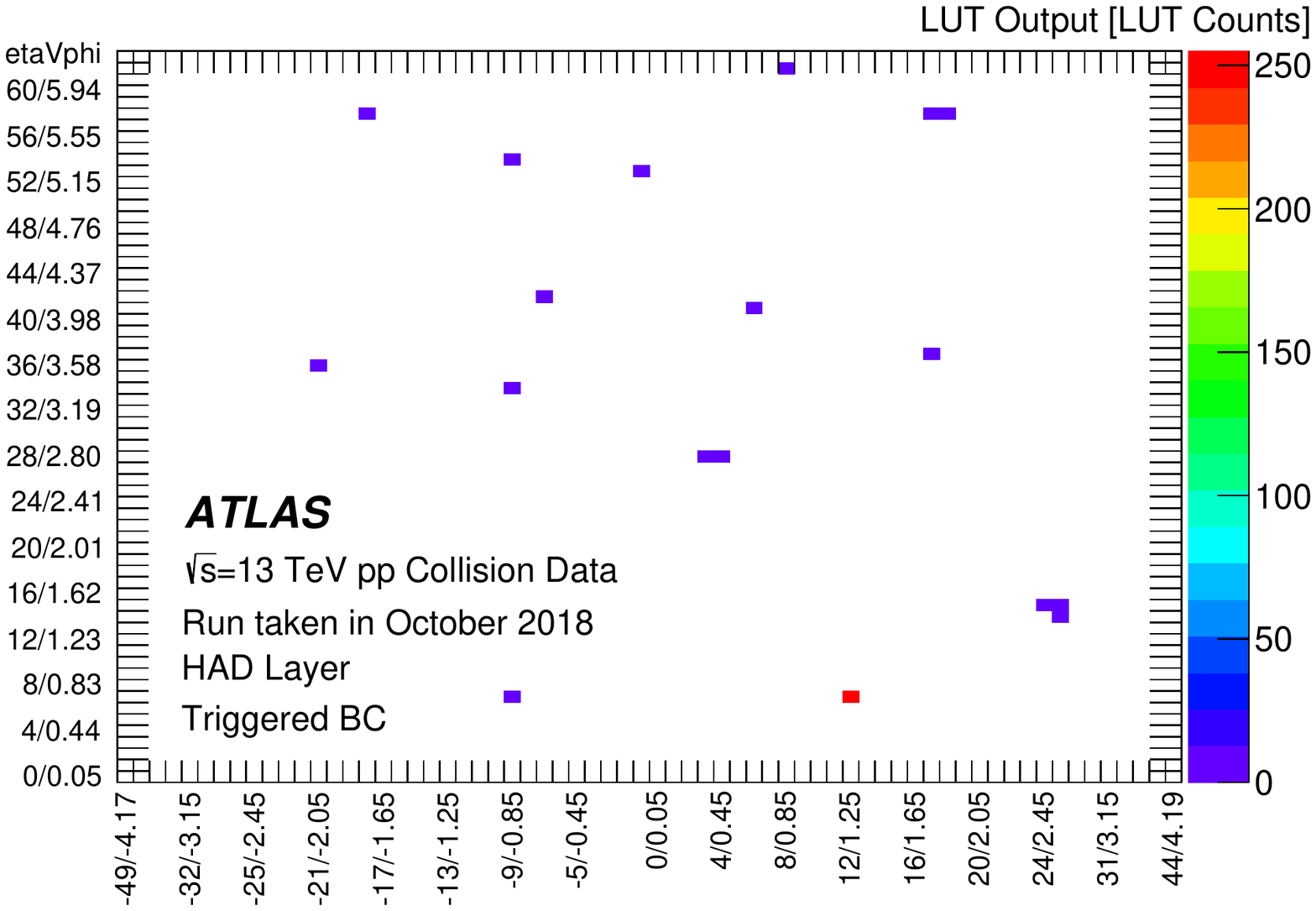}
\caption[]{}
\end{subfigure}
\begin{subfigure}[b]{0.82\textwidth}
\centering
\includegraphics[width=\textwidth]{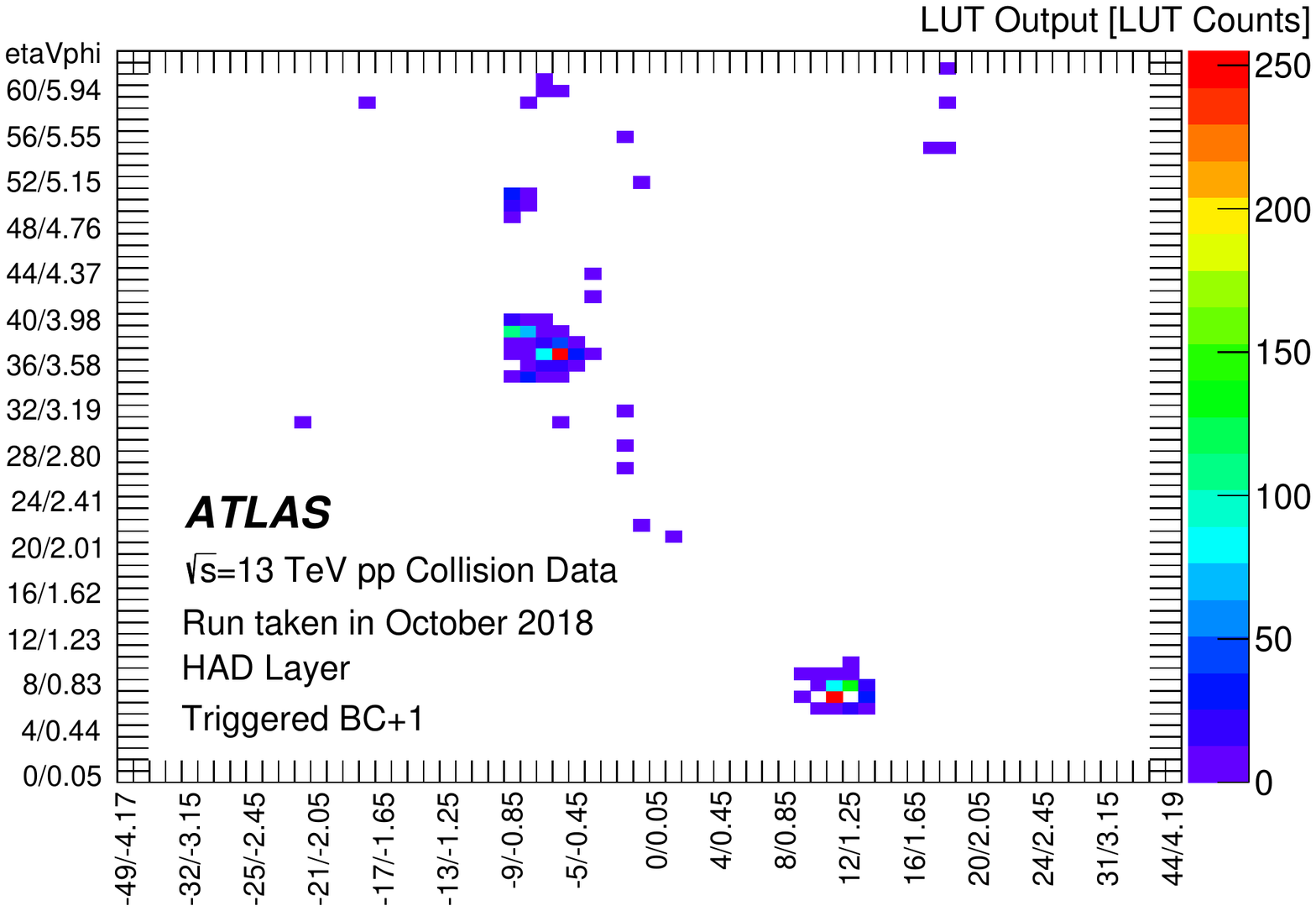}
\caption[]{}
\end{subfigure}
\caption[Early Triggered Event Example]{Example of an event triggered one BC early by the L1Calo BCID algorithm.
The figures show $\eta$--$\phi$ maps of the LUT output for all TTs in the hadronic layer for two consecutive BCs.
The axis labels indicate both the real $\eta$ and $\phi$ coordinates as well as the corresponding integer bin numbers (cf.\ Section~\ref{sec:inputpath}).
In the hadronic layer, one~LUT count corresponds to approximately \SI{1}{\giga\electronvolt} (cf.\ Section~\ref{sec:performance_lut}).
The plots are taken from the automated monitoring of the Saturated BCID performance.}
\label{fig:sat80monitoring}
\end{figure}

Figure~\ref{fig:sat80monitoring} shows an example of an early triggered event, as seen in the 2018 monitoring.
The two plots show, for two consecutive BCs, an $\eta$--$\phi$ map of the LUT output of all TTs in the hadronic calorimeter.
Figure~\ref{fig:sat80monitoring}(b) shows the result for the BC after the triggered BC, and contains two large clusters of TTs, corresponding to a high-energy dijet event.
Figure~\ref{fig:sat80monitoring}(a) shows the result for the triggered BC, which only contains one TT with a significant energy deposition.
 
\subsection{CP \& JEP Look-Up Tables}
\label{sec:performance_lut}
 
Before entering the PPr, the analogue TT signals are routed through the Receiver system, which includes the initial
$E_{\textrm{T}}$ calibration of all signals: gain factors are set such that one ADC count in the PPr corresponds
to \SI{250}{\mega\electronvolt} (see Section~\ref{sec:inputpath}).
The final $E_{\textrm{T}}$ calibration is then performed in a Look-Up Table (LUT) in the MCM, which also includes
pedestal subtraction, noise suppression, and turning off problematic channels.
 
The output of the MCM digital filter is converted to \si{\giga\electronvolt} before it is transmitted to the CP and JEP systems.
Each of the possible 16-bit filter results is assigned an 8-bit energy value in a two-step procedure:
\begin{itemize}
\item The filter output is reduced to 10 bits by dropping the least significant 6 bits.
\item The reduced filter output is passed through a LUT that assigns an 8-bit value to each 10-bit input.
\end{itemize}
 
For the first step, bits are dropped in order to achieve the same normalisation for all pulses, as they can differ chosen filter coefficients.
 
For the second step, the LUT content typically resembles a linear function, saturating at $E_{\textrm{T}}$ outputs of \SI{255}{\giga\electronvolt}, with a resolution of \SI{1}{\giga\electronvolt} per least significant bit (LSB).
Additionally, the LUT is used to implement thresholds for the signal pedestal and the noise, by applying an offset to the linear function and assigning non-zero $E_{\textrm{T}}$ only to filter outputs above a configurable noise threshold.
Following this approach, the number of fake $E_{\textrm{T}}$ assignments due to statistical fluctuations in the electronic or pile-up noise can be controlled in a systematic way.
By permanently setting the LUT content to zero, problematic channels can be turned off entirely.
 
In contrast to the MCM, in which only one LUT per TT was available, the nMCM uses two independently configurable LUTs.
This allows the $E_{\textrm{T}}$ output to be calibrated separately for the CP and JEP systems, to optimise the performance of each system without impacting the other.
 
In the case of the CP system, the accuracy of isolation criteria and hadronic vetoes is improved by doubling the resolution to \SI{0.5}{\giga\electronvolt} per LSB.
The resulting lower LUT saturation at \SI{127.5}{\giga\electronvolt} per TT does not impact the L1Calo Trigger performance, as all electromagnetic trigger items are below this energy threshold. The evolution and performance of the EM L1 triggers during Run~2 is discussed in detail in Ref.~\cite{egammatrig}.
 
The JEP-LUT continues to use the standard configuration with a resolution of \SI{1}{\giga\electronvolt}.
A non-linear configuration that allows lower noise cuts to be set was studied, and was shown to improve both the jet trigger efficiency and the $E_{\textrm{T}}^{\textrm{miss}}$ trigger purity.
However, this configuration was also found to be very sensitive to the exact filling conditions of the LHC, effectively requiring extensive recalibrations for even slight changes in these conditions.
This approach was thus abandoned in favour of keeping the known, stable and more maintainable configuration.
 
The chosen noise thresholds in the two LUTs play an essential role in determining the efficiency and purity of the L1Calo Trigger items.
For the CP system, which identifies narrow clusters in only the central $\eta$ region of the detector, fake $E_{\textrm{T}}$ assignments mainly alter the performance of the electromagnetic and hadronic isolation requirements.
Hence, low cut values of approximately \SI{1}{\giga\electronvolt} were used in the CP-LUT for all trigger towers throughout Run~2.
This corresponds to 2--3 times the ADC noise in the absence of input signals for the towers in the EM and HAD barrels, respectively (cf.\ Figure~\ref{fig:NoiseFullPpr}).
 
For the large objects identified by the JEP system in the full detector acceptance, statistical fluctuations in the pile-up noise have a strong impact on the purity, and thus rate, of the corresponding trigger items. This mainly affects the $E_{\textrm{T}}^{\textrm{miss}}$ triggers, for which more details can be found in~\cite{Aad:2020les}.
Hence, the noise cuts in the JEP-LUT are set according to the occupancy of the TTs as a function of $\eta$, by assigning only the highest 0.5\% of filter outputs within each $\eta$-bin to non-zero $E_{\textrm{T}}$ values~\cite{ad_thesis}.
Since the TT occupancy depends on the LHC operating conditions, the noise cuts have to be recalibrated in case of major changes, e.g.\ in the beam filling scheme, proton bunch intensity and brilliance.
 
\begin{figure}[t]
\centering
\begin{subfigure}[b]{0.49\textwidth}
\centering
\includegraphics[width=\textwidth]{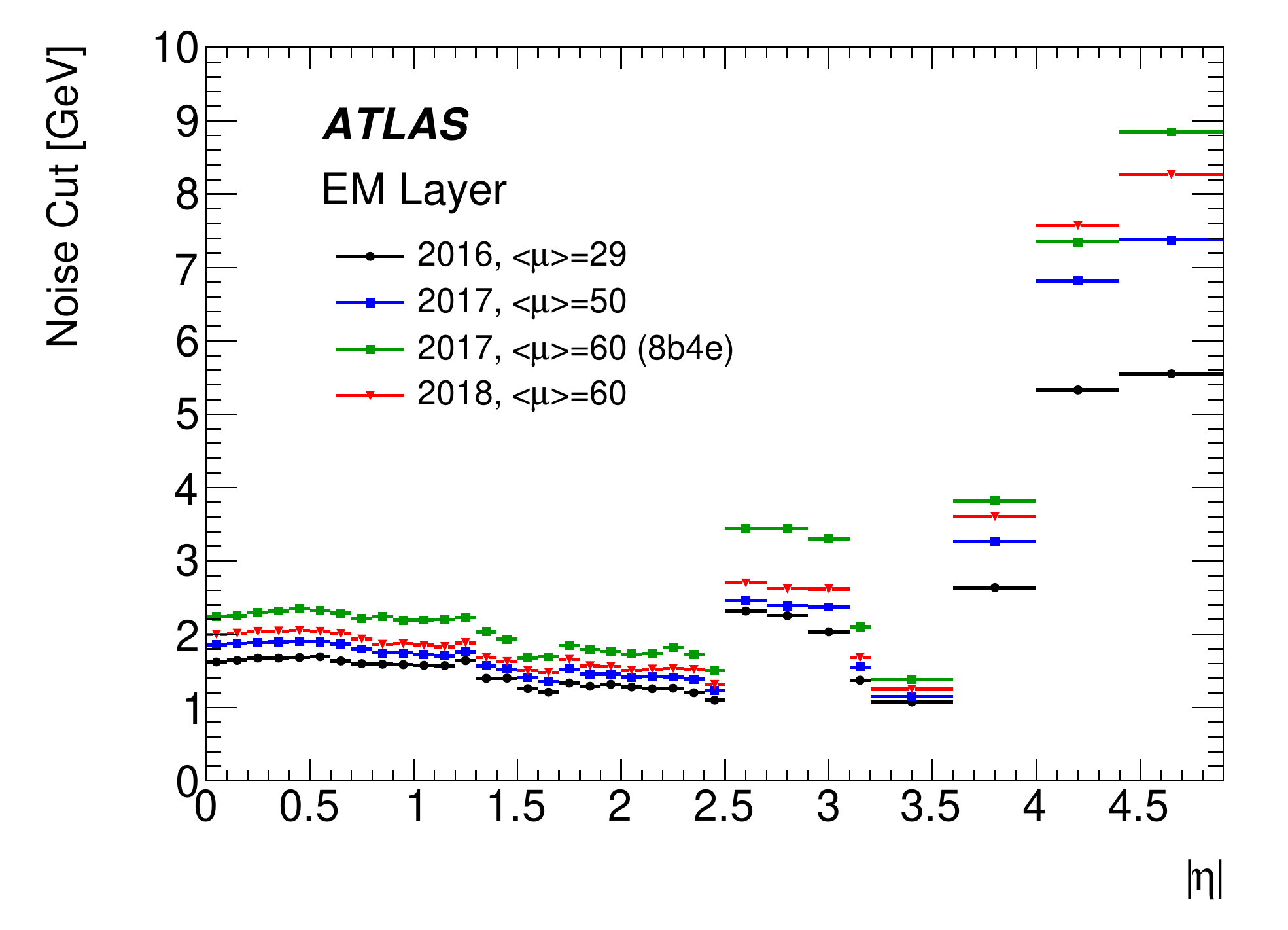}
\caption[]{}
\end{subfigure}
~
\begin{subfigure}[b]{0.49\textwidth}
\centering
\includegraphics[width=\textwidth]{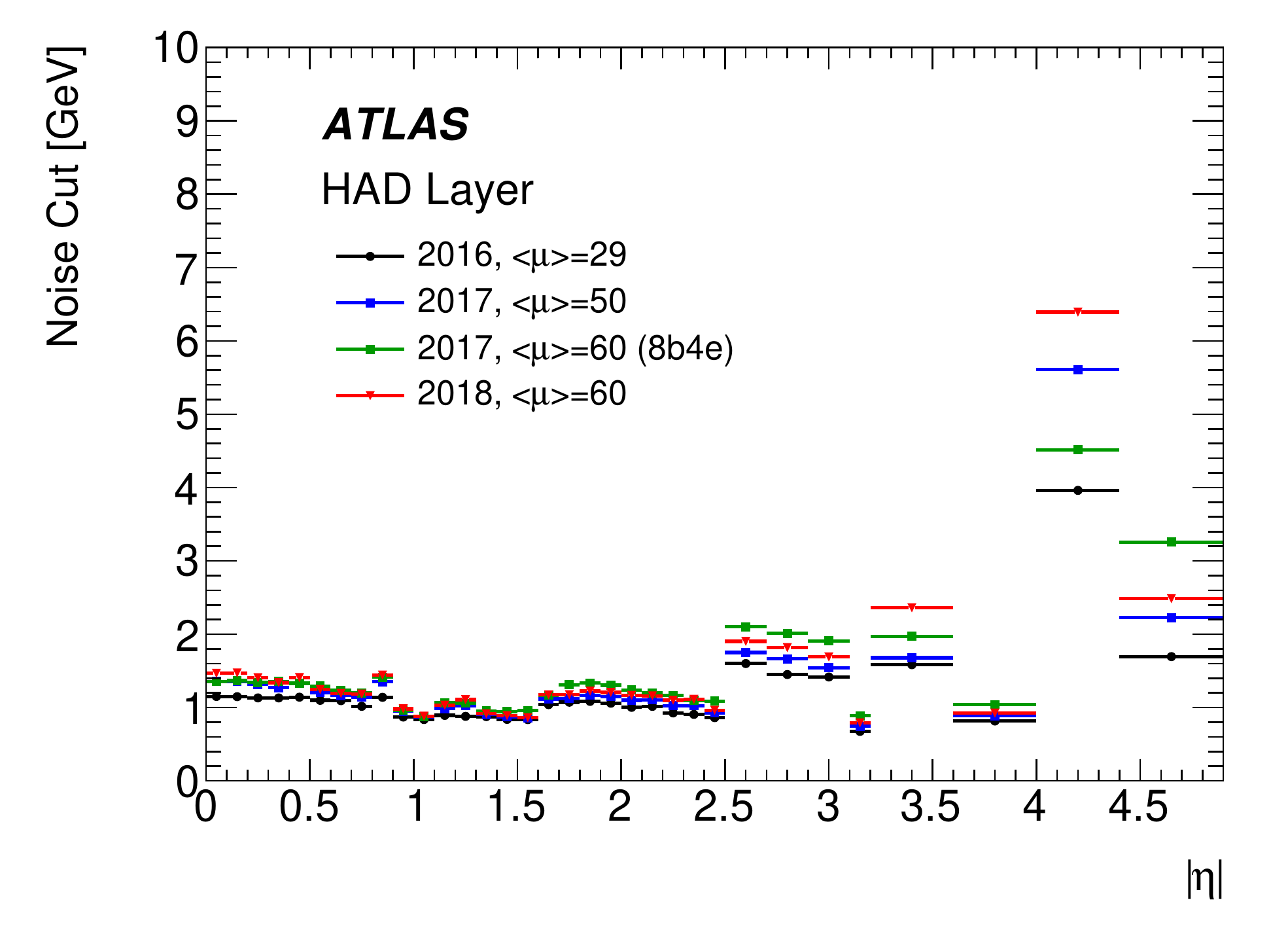}
\caption[]{}
\end{subfigure}
\caption[JEP-LUT noise thresholds used during Run~2]
{JEP-LUT noise thresholds used during Run~2 for (a) the EM layer and (b) the HAD layer as function of $|\eta|$,
separately for individual data-taking periods. The given $\langle\mu\rangle$ values correspond to peak average numbers of $pp$ interactions per bunch crossing.
Artificial $|\eta|$ values (3.4, 3.8, 4.2, 4.65) have been assigned to the four hadronic FCAL towers, which in reality cover twice the $\eta$ range
but are situated behind each other. }
\label{fig:nosiecuts}
\end{figure}
 
The JEP-LUT noise thresholds which were used for most of Run~2 are shown in Figure~\ref{fig:nosiecuts} separately for different data-taking periods.
They range from an $E_{\textrm{T}}$ of 1--\SI{2}{\giga\electronvolt} in the central $\eta$ region to as high as \SI{10}{\giga\electronvolt} in the forward $\eta$ region.
The noise cuts are tuned for the peak numbers of $pp$ interactions per bunch crossing reached in the individual data-taking periods,
and increase with increasing pile-up, most significantly in the wide FCAL1 towers which are located closest to the beam.
Overall, lower cuts are applied in the hadronic layer owing to the reduced impact of pile-up due to the shielding provided by the EM calorimeter.
The worsened pile-up conditions during LHC operation with the 8b4e filling scheme in the second half of 2017 led to the highest noise cuts used during Run~2,
for the most part exceeding those in 2018 with identical peak $\langle\mu\rangle$.

\FloatBarrier
 
\section{Summary}
\label{sec:summary}
The increased centre-of-mass energy and luminosity in Run~2 was expected to pose challenges to the LHC experiments, especially to the trigger and readout electronics.
With a value of $1.9\times 10^{34}$\,cm$^{-2}$\,s$^{-1}$ the instantaneous luminosity reached almost twice the nominal value causing peak numbers of proton--proton interactions per bunch crossing of more than 60, which exceed the LHC design value of 27.
 
To continue operating with as high efficiency as achieved during Run~1, the L1Calo Trigger of the ATLAS experiment was upgraded during the first long shutdown of the LHC.
In the L1Calo PreProcessor the MCMs were replaced by nMCMs, featuring modern FADCs and FPGA-based digital signal processing.
 
By means of new hardware implementation, many updates in the signal processing were successfully implemented and operated during Run~2, improving the performance of the L1Calo Trigger.
To this end, the digital filter implementation was expanded to allow autocorrelation filter coefficients to be used, which ensures a high efficiency of the Peak-Finder bunch crossing identification algorithm even in high pile-up scenarios.
A dynamic pedestal correction algorithm was implemented, which calculates and corrects for the average pile-up contribution individually for each trigger tower and for each bunch in the LHC orbit, lowering fake $E_{\textrm{T}}^{\textrm{miss}}$ trigger rates substantially.
The treatment of signals which saturate the FADCs was also improved using the higher sampling rate provided by the FADCs on the nMCM, such that signals up to the highest possible energies are assigned to the correct bunch crossing.
With all these improvements, the PreProcessor contributed significantly to the successful operation of the L1Calo Trigger in Run~2.
 
The PreProcessor along with the whole Run~2 L1Calo system will be in operation for the start of Run~3, providing data to aid the commissioning of the future L1Calo Trigger.
After this initial phase, the PreProcessor crates receiving data from the Tile calorimeter will continue to operate for the whole duration of Run~3, providing digital transverse energy results to the upgraded L1Calo Trigger.

\section*{Acknowledgements}
 

We thank CERN for the very successful operation of the LHC, as well as the
support staff from our institutions without whom ATLAS could not be
operated efficiently.
 
We acknowledge the support of ANPCyT, Argentina; YerPhI, Armenia; ARC, Australia; BMWFW and FWF, Austria; ANAS, Azerbaijan; SSTC, Belarus; CNPq and FAPESP, Brazil; NSERC, NRC and CFI, Canada; CERN; ANID, Chile; CAS, MOST and NSFC, China; COLCIENCIAS, Colombia; MSMT CR, MPO CR and VSC CR, Czech Republic; DNRF and DNSRC, Denmark; IN2P3-CNRS and CEA-DRF/IRFU, France; SRNSFG, Georgia; BMBF, HGF and MPG, Germany; GSRT, Greece; RGC and Hong Kong SAR, China; ISF and Benoziyo Center, Israel; INFN, Italy; MEXT and JSPS, Japan; CNRST, Morocco; NWO, Netherlands; RCN, Norway; MNiSW and NCN, Poland; FCT, Portugal; MNE/IFA, Romania; MES of Russia and NRC KI, Russia Federation; JINR; MESTD, Serbia; MSSR, Slovakia; ARRS and MIZ\v{S}, Slovenia; DST/NRF, South Africa; MICINN, Spain; SRC and Wallenberg Foundation, Sweden; SERI, SNSF and Cantons of Bern and Geneva, Switzerland; MOST, Taiwan; TAEK, Turkey; STFC, United Kingdom; DOE and NSF, United States of America. In addition, individual groups and members have received support from BCKDF, CANARIE, Compute Canada and CRC, Canada; ERC, ERDF, Horizon 2020, Marie Sk{\l}odowska-Curie Actions and COST, European Union; Investissements d'Avenir Labex, Investissements d'Avenir Idex and ANR, France; DFG and AvH Foundation, Germany; Herakleitos, Thales and Aristeia programmes co-financed by EU-ESF and the Greek NSRF, Greece; BSF-NSF and GIF, Israel; La Caixa Banking Foundation, CERCA Programme Generalitat de Catalunya and PROMETEO and GenT Programmes Generalitat Valenciana, Spain; G\"{o}ran Gustafssons Stiftelse, Sweden; The Royal Society and Leverhulme Trust, United Kingdom.
 
The crucial computing support from all WLCG partners is acknowledged gratefully, in particular from CERN, the ATLAS Tier-1 facilities at TRIUMF (Canada), NDGF (Denmark, Norway, Sweden), CC-IN2P3 (France), KIT/GridKA (Germany), INFN-CNAF (Italy), NL-T1 (Netherlands), PIC (Spain), ASGC (Taiwan), RAL (UK) and BNL (USA), the Tier-2 facilities worldwide and large non-WLCG resource providers. Major contributors of computing resources are listed in Ref.~\cite{ATL-SOFT-PUB-2020-001}.
 

\printbibliography

\clearpage
 
\begin{flushleft}
\hypersetup{urlcolor=black}
{\Large The ATLAS Collaboration}

\bigskip

\AtlasOrcid[0000-0002-6665-4934]{G.~Aad}$^\textrm{\scriptsize 102}$,    
\AtlasOrcid[0000-0002-5888-2734]{B.~Abbott}$^\textrm{\scriptsize 128}$,    
\AtlasOrcid{D.C.~Abbott}$^\textrm{\scriptsize 103}$,    
\AtlasOrcid[0000-0002-2788-3822]{A.~Abed~Abud}$^\textrm{\scriptsize 36}$,    
\AtlasOrcid[0000-0002-1002-1652]{K.~Abeling}$^\textrm{\scriptsize 53}$,    
\AtlasOrcid[0000-0002-2987-4006]{D.K.~Abhayasinghe}$^\textrm{\scriptsize 94}$,    
\AtlasOrcid[0000-0002-8496-9294]{S.H.~Abidi}$^\textrm{\scriptsize 166}$,    
\AtlasOrcid[0000-0002-8279-9324]{O.S.~AbouZeid}$^\textrm{\scriptsize 40}$,    
\AtlasOrcid{N.L.~Abraham}$^\textrm{\scriptsize 155}$,    
\AtlasOrcid[0000-0001-5329-6640]{H.~Abramowicz}$^\textrm{\scriptsize 160}$
% ...
\bigskip
\\

% ...
$^{36}$CERN, Geneva; Switzerland.\\
% ...
$^{40}$Niels Bohr Institute, University of Copenhagen, Copenhagen; Denmark.\\
% ...
$^{53}$II. Physikalisches Institut, Georg-August-Universit\"{a}t G\"ottingen, G\"ottingen; Germany.\\
% ...
$^{94}$Department of Physics, Royal Holloway University of London, Egham; United Kingdom.\\
% ...
$^{102}$CPPM, Aix-Marseille Universit\'e, CNRS/IN2P3, Marseille; France.\\
$^{103}$Department of Physics, University of Massachusetts, Amherst MA; United States of America.\\
% ...
$^{128}$Homer L. Dodge Department of Physics and Astronomy, University of Oklahoma, Norman OK; United States of America.\\
% ...
$^{155}$Department of Physics and Astronomy, University of Sussex, Brighton; United Kingdom.\\
% ...
$^{160}$Raymond and Beverly Sackler School of Physics and Astronomy, Tel Aviv University, Tel Aviv; Israel.\\
% ...
$^{166}$Department of Physics, University of Toronto, Toronto ON; Canada.\\
% ...

\end{flushleft}

 
\end{document}